\documentclass[conference]{IEEEtran}
\usepackage{nicefrac}
\usepackage{siunitx}
\usepackage{array,framed}%array用来修改表格行高
\usepackage{booktabs}
\usepackage{
  color,
  float,
  epsfig,
  wrapfig,
  graphics,
  graphicx,
  subcaption
}
\usepackage{ulem,CJKulem}
\usepackage{textcomp,amssymb}
\usepackage{setspace}
\usepackage{latexsym,fancyhdr,url}
\usepackage{enumerate}
\usepackage{algorithm2e}
\usepackage{algpseudocode}
\usepackage{graphics}
\usepackage{xparse} % argument parsing -- \edist
\usepackage{xspace}
\usepackage{multirow}
\usepackage{csvsimple}
\usepackage{balance}
\usepackage{amsmath}
\usepackage{xcolor}
\usepackage{pifont}
\usepackage{CJKutf8}
\usepackage{
  tikz,
  pgfplots,
  pgfplotstable
}
\usepackage{hyperref}

\usetikzlibrary{
  shapes.geometric,
  arrows,
  external,
  pgfplots.groupplots,
  matrix
}
\usepackage{colortbl}

\usepackage{xparse}
\usepackage{listings}
\newcommand{\bnm}{\begin{newmath}}
\newcommand{\enm}{\end{newmath}}

\newcommand{\bea}{\begin{eqnarray*}}%
\newcommand{\eea}{\end{eqnarray*}}%

\newcommand{\bne}{\begin{newequation}}
\newcommand{\ene}{\end{newequation}}

\newcommand{\bal}{\begin{newalign}}
\newcommand{\eal}{\end{newalign}}

\newenvironment{newalign}{\begin{align}%
\setlength{\abovedisplayskip}{4pt}%
\setlength{\belowdisplayskip}{4pt}%
\setlength{\abovedisplayshortskip}{6pt}%
\setlength{\belowdisplayshortskip}{6pt} }{\end{align}}

\newenvironment{newmath}{\begin{displaymath}%
\setlength{\abovedisplayskip}{4pt}%
\setlength{\belowdisplayskip}{4pt}%
\setlength{\abovedisplayshortskip}{6pt}%
\setlength{\belowdisplayshortskip}{6pt} }{\end{displaymath}}

\newenvironment{newequation}{\begin{equation}%
\setlength{\abovedisplayskip}{4pt}%
\setlength{\belowdisplayskip}{4pt}%
\setlength{\abovedisplayshortskip}{6pt}%
\setlength{\belowdisplayshortskip}{6pt} }{\end{equation}}

\newcounter{ctr}

\newcounter{mytable}
\def\mytable{\begin{centering}\refstepcounter{mytable}}
\def\endmytable{\end{centering}}

\newcounter{myfig}
\def\myfig{\begin{centering}\refstepcounter{myfig}}
\def\endmyfig{\end{centering}}

\newlength{\saveparindent}
\newlength{\saveparskip}
\newcommand{\E}{{\rm I\kern-.3em E}}

\renewcommand{\eqref}[1]{\mbox{Equation~(\ref{#1})}}

\def \part {part}

\renewcommand{\paragraph}[1]{\vspace*{6pt}\noindent\textbf{#1}\;}

\def \blackslug{\hbox{\hskip 1pt \vrule width 4pt height 8pt
    depth 1.5pt \hskip 1pt}}
\def \qed{\quad\blackslug\lower 8.5pt\null\par}
\newcounter{mynote}[section]

\newcommand{\todo}[1]{{\textcolor{green}{[TODO: #1]}}}

\newcounter{rcnote}[section]

\newcounter{mrnote}[section]

\newcounter{fknote}[section]

\newcounter{anote}[section]

\DeclareMathSymbol{\mlq}{\mathord}{operators}{``}
\DeclareMathSymbol{\mrq}{\mathord}{operators}{`'}

\newcommand{\rhf}[2]{R_{f, \gamma}}

\DeclareDocumentCommand{\edist}{o o}{
  \ensuremath{
    \IfNoValueTF{#1}{{d}}{{\sf d}(#1,#2)}
  }
}

\newcommand{\olrk}[1]{\ifx\nursymbol#1\else\!\!\mskip4.5mu plus 0.5mu\left(\mskip0.5mu plus0.5mu #1\mskip1.5mu plus0.5mu \right)\fi}

\NewDocumentCommand{\indseq}{ O{1} O{r} }{{#1}\ldots {#2}}

\usepackage{tikz}

\usepackage{threeparttable}

\definecolor{darkgreen}{HTML}{24A10B}

\newcommand{\ignore}[1]{}

\newcommand{\td}[1]{\textcolor{black}{#1}}
\newcommand{\revise}[1]{\textcolor{black}{#1}}
\newcommand{\yy}[1]{\textcolor{black}{#1}}
\newcommand{\nw}[1]{\textcolor{black}{#1}}

\ifCLASSINFOpdf
\else
\fi
\newcommand{\chatd}[1]{\textit{ChatDetector}}
\newcommand{\chat}[1]{ChatGPT}
\newcommand{\llm}[1]{LLMs}

\begin{document}
\title{The Midas Touch: Triggering the Capability of LLMs for RM-API Misuse Detection\\}


\IEEEoverridecommandlockouts
\makeatletter\def\@IEEEpubidpullup{6.5\baselineskip}\makeatother
\IEEEpubid{\parbox{\columnwidth}{
		Network and Distributed System Security (NDSS) Symposium 2025\\
		23-28 February 2025, San Diego, CA, USA\\
		ISBN 979-8-9894372-8-3\\
	  https://dx.doi.org/10.14722/ndss.2025.23816\\
		www.ndss-symposium.org
}
\hspace{\columnsep}\makebox[\columnwidth]{}}

\author{
\IEEEauthorblockN{Yi Yang$^{1,2}$, Jinghua Liu$^{1,2}$, Kai Chen$^{1,2,\star}$\thanks{$\star$\ \text{Corresponding Author}}, and Miaoqian Lin$^{1,2}$}\\
\IEEEauthorblockA{
$^1$Institute of Information Engineering, Chinese Academy of Sciences, China\\
$^2$School of Cyber Security, University of Chinese Academy of Sciences, China\\
{\{yangyi, liujinghua, chenkai, linmiaoqian\}@iie.ac.cn}}}

\maketitle
%\fancyhf{} 
%\fancyfoot[C]{\thepage}



% \fi
%%\IEEEpeerreviewmaketitle

% Section I
\begin{abstract}

As the basis of software resource management (RM), strictly following the RM-API constraints guarantees secure resource management and software.
To enhance the RM-API application, researchers find it effective in detecting RM-API misuse on open-source software according to RM-API constraints retrieved from documentation and code.   
However, the current pattern-matching constraint retrieval methods have limitations: the documentation-based methods leave many API constraints irregularly distributed or involving neutral sentiment undiscovered; the code-based methods result in many false bugs due to incorrect API usage since not all high-frequency usages are correct.
Therefore, people propose to utilize Large Language Models (LLMs) for RM-API constraint retrieval with their potential on text analysis and generation.
However, directly using LLMs has limitations due to the hallucinations.
The LLMs fabricate answers without expertise leaving many RM APIs undiscovered and generating incorrect answers even with evidence introducing incorrect RM-API constraints and false bugs.

In this paper, we propose an LLM-empowered RM-API misuse detection solution, \chatd{}, which fully automates \llm{} for documentation understanding which helps RM-API constraints retrieval and RM-API misuse detection.
To correctly retrieve the RM-API constraints, \chatd{} is inspired by the ReAct framework which is optimized based on Chain-of-Thought (CoT) to decompose the complex task into allocation APIs identification, RM-object (allocated/released by RM APIs) extraction and RM-APIs pairing (RM APIs usually exist in pairs). 
It first verifies the semantics of allocation APIs based on the retrieved RM sentences from API documentation through \llm{}. 
Inspired by the \llm{}' performance on various prompting methods, \chatd{} adopts a two-dimensional prompting approach for cross-validation.
At the same time, an inconsistency-checking approach between the \llm{}' output and the reasoning process is adopted for the allocation APIs confirmation with an off-the-shelf Natural Language Processing (NLP) tool.
To accurately pair the RM-APIs, \chatd{} decomposes the task again and identifies the RM-object type first, with which it can then accurately pair the releasing APIs and further construct the RM-API constraints for misuse detection.  
With the diminished hallucinations, \chatd{} identifies 165 pairs of RM-APIs with a precision of 98.21\% compared with the state-of-the-art API detectors.
By employing a static detector CodeQL, we ethically report 115 security bugs on the applications integrating on six popular libraries to the developers, which may result in severe issues, such as Denial-of-Services (DoS) and memory corruption.
Compared with the end-to-end benchmark method, the result shows that \chatd{} can retrieve at least 47\% more RM sentences and 80.85\% more RM-API constraints.
Since no work exists specified in utilizing \llm{} for RM-API misuse detection to our best knowledge, the inspiring results show that \llm{} can assist in generating more constraints beyond expertise and can be used for bug detection. It also indicates that future research could transfer from overcoming the bottlenecks of traditional NLP tools to creatively utilizing LLMs for security research.

\end{abstract}
\section{Introduction}
\label{sec:intro}

%background
%\todo{background+motivations}
Creating applications requires utilizing Application Programming Interfaces (APIs), and developers are expected to refer to library documentation to ensure their appropriate implementation. Safely utilizing resource-management (RM) APIs, usually employed in pairs (for instance, allocation and releasing), demands \revise{strictly} adherence to explicitly outlined constraints. Breaching these constraints could lead to security vulnerabilities, including but not limited to memory leaks, use-after-free issues, and double-free security bugs.

\revise{
Traditional methods use common-used keywords to identify RM-APIs~\cite{zhong2011inferring}.}
Recently many other studies have proposed to achieve API misuse detection by analyzing API documentation~\cite{lv2020rtfm,pandita2012inferring} within which the API constraints are described in strong sentiment or by detecting deviations from frequently used API usages which could be extracted from code analysis~\cite{lyu2022goshawk,wu2021understanding,yun2016apisan}.
However, all these methods are limited by expertise.
\revise{
Directly using keywords will result in large number of false negatives, the statistics say it will introduce at least 76.28\% RM-APIs missing from identification.}
For those involving documentation analysis, it is hard to retrieve the API constraints beyond the documentation; 
For those focusing on code analysis, the incorrect usage will influence the whole detection procedure introducing false bugs. 
\revise{Traditional program analysis also has difficulties identifying if the current allocation function needs extra releasing operations within multi-layered nesting. The statistics of the GroundTruth in Section~\ref{sec:explore} \yy{shows} 72.7\% of functions have multi-layered nesting and merely 27.3\% of functions directly \yy{call} \texttt{malloc}/\texttt{free} functions. These results show that traditional program analysis will introduce false negatives in RM-API pairing due to complex code structures; it also introduces false positives when the allocated/released object has already been released within the function and no extra releasing operations are required.}
Therefore, with the potential of Large Language Models (LLMs) on text analysis and generation, researchers tend to explore the ability of LLMs on bug detection.

%challenge
\noindent\textbf{Challenges in exploiting LLMs for RM-API misuse detection}.
We attempt to use one of the state-of-the-art LLMs to retrieve information from API official documentation, identify resource allocation functions, and pair them with the corresponding releasing functions. However, directly using LLMs still encounters the following challenges.

\noindent\textbf{C1: LLMs fabricate answers without expertise.}
We observed that when directly prompting \llm{} to provide corresponding releasing functions based on the existing allocation functions, \llm{} tend to fabricate answers according to the naming format.
For example, \texttt{evwatch\_check\_new} and \texttt{evwatch\_free} form a pair of RM-APIs, but \llm{} provide the resource-releasing function \texttt{evwatch\_check\_free}, which follows a naming pattern similar to \texttt{evwatch\_check\_new}, and the operation part includes a typical antonym pair, i.e., \textit{new} and \textit{free}. However, this function does not exist, and \llm{} exhibit obvious fabrication probably due to the absence of information. Therefore, supplementing valid information and designing the solution procedure are crucial in addressing the fabrication issues.

\noindent\textbf{C2: \llm{} introduce incorrect answers with evidence.}
Another challenge arises from \llm{} producing incorrect answers even with grounded evidence. As a black-box model, When identifying the current function's semantics based on specific sentences with \llm{}, we cannot guide the model to output the expected results.
For example, after prompting \llm{} with the RM sentence \textit{``The zip\_discard() function closes the archive and frees the memory allocated for it''} of \texttt{zip\_discard}, and prompting \llm{} to determine if the current function has allocation semantics. Even though the RM sentence contains the keyword \textit{free} indicating an obvious resource-releasing semantics, \llm{} still provide the incorrect answer \textit{YES}. This phenomenon is quite common and can impact the final identification of allocation functions, potentially leading to false bugs. Therefore, solving the encountered issues in \llm{} and generating accurate answers is also important.

%main problem is incorrect answers, two reasons leading to this problem
%\textbf{\textit{C1: Missing information.}}

%\textbf{\textit{C2: Incorrect information.}}

%describe the tool
\noindent\textbf{\chatd{}.}
%\todo{Revise}
To address the challenges mentioned above, we propose an LLM-empowered RM-API misuse detection solution, \chatd{}.
\nw{The challenges mentioned above happen in most of the state-of-the-art LLMs (see Section~\ref{sec:discuss}) and we perform \chatd{} based on \chat{} considering the overall performance.}
We adopt the inspiration from ReACT framework~\cite{yao2023react} and compose \chatd{} including three components: 1) utilizing \chat{} to identify resource allocation functions from API documentation; 2) employing \chat{} to provide corresponding resource-releasing functions for the identified resource allocation functions; 3) using CodeQL together with the identified RM-API pairs to detect misuses in open-source software.

\ignore{Specifically, we observed that the reason \llm{} randomly fabricate answers is due to hallucinations~\cite{ji2023survey}. And the presence of excessive irrelevant information in the input text, causes \llm{} to exhibit a \textit{guessing} behaviour when answering questions. To address the problems above, we propose a method that involves using \llm{} to extract relevant knowledge from API official documentation and assist in resource-allocation functions identification.}

First, \chatd{} decomposes the complex task into two steps: identifying RM (Resource-management) sentences from API official documentation that describe the current function's functionality and are related to resource-allocation operations; and then, determining which function has resource allocation semantics based on the RM sentences.
Specifically, the recognized RM sentences come in two types: 1) RM sentences that contain the function name and the allocated object (RM object); and 2) RM sentences that only include the allocated object.
According to the various prompting methods and information, we design a two-dimensional prompting method to get the reasoning answers for allocation APIs identification. After that, \chatd{} employs cross-validation for the \llm{}' output to identify the potential allocation APIs.
Through result inspection, we discover that the current method still prompts \llm{} to produce evident errors in recognizing function semantics based on RM sentences. For example, when RM sentences contain obvious keywords indicating resource-releasing semantics like \textit{free}, \chat{} still erroneously identifies the current function's semantics as resource allocation. This obvious inconsistency between the input and output results in a significant number of false positives.
To address this issue, we propose an inconsistency check between the \llm{}'s output and the reasoning process. Specifically, when the current function is identified as having allocation semantics, \chatd{} uses Semantic Role Labeling~\cite{semantic} (an off-the-shelf NLP tool) to extract the operation keywords corresponding to the current function during the reasoning process. Then \chatd{} will classify the operation into \textit{allocate}, \textit{free} or \textit{neutral} to further validate that the current answer is generated based on the original text rather than fabricated.
Through cross-validation of answers provided after the three prompts and inconsistency checks on the reasoning process, \chatd{} can recognize resource allocation functions and their corresponding descriptions of RM objects.

Then, \chatd{} attempts to prompt \llm{} to provide the corresponding resource-releasing functions based on the identified resource-allocation functions. 
Due to the same issues encountered above, \llm{} also tend to fabricate answers based on the function's naming format due to incomplete information and complex tasks.
Therefore, based on the fact that both resource-management functions act on the same RM object, \chatd{} decomposes the pairing task, which is inspired by the ReAct framework, into RM-object type extraction and releasing API pairing. 
Specifically, \chatd{} locates the corresponding RM-object type in the current function declaration based on the RM object description. 
Due to the poor performance of \llm{} in identifying object types based on unstructured and lengthy function declarations, \chatd{} pre-processes the function declarations into formatted data, distinguishing the return type and name as well as the types and names of each parameter, which enables \llm{} to better identify RM-object types. Based on the identified RM object types, \chatd{} can generate accurate resource-releasing functions with diminished ``hallucinations'' in \llm{}. This illustrates that the current approach, combining external knowledge and step-by-step reasoning, can enhance the accuracy of \llm{} in identifying RM-API pairs.

Finally, we evaluate \chatd{} using the following research questions:

\textbf{RQ1:} Can \chatd{} be utilized for RM-API misuse detection?
%\textbf{RQ1: Can \chatd{} be utilized for RM-API misuse detection?}
\textbf{RQ2:} Can \chatd{} discover more RM-API pairs than the previous work?
%\textbf{RQ2: Can \chatd{} discover more RM-API pairs than the previous work?}
\textbf{RQ3:} Is it necessary to perform reasoning with \llm{} for RM-API misuse detection?
%\textbf{RQ3: Is it necessary to perform reasoning with LLMs for RM-API misuse detection?}

On the dataset including six libraries (FFmpeg~\cite{Ffmpeg}, Libevent~\cite{libevent}, Libexpat~\cite{libexpat}, Libpcap~\cite{Libpcap}, Libzip~\cite{Libzip} and OpenLdap~\cite{Openldap}), \chatd{} can identify 165 RM-API pairs, with relatively high precision of 92.81\% in allocation APIs identification and 98.21\% in RM-API pairing, outperforming the benchmark work by identifying 80.85\% more RM-API pairs. 
%On a larger dataset of applications integrated on 6 popular libraries , 
\chatd{} discovers 115 security issues which were ethically reported to the developers (in Section~\ref{subsuc:eval}).

%discovery
%\paragraph{Discoveries.}

\noindent\textbf{Contributions.}
Our contributions are listed as follows:

\vspace {1pt}\noindent$\bullet$\space\textbf{\textit{Novel research direction.}}
To the best of our knowledge, \chatd{} is the first to propose using \llm{} to understand and extract API usage patterns from official API documentation, enabling misuse detection in real software.
This work opens up new research directions for bug detection aided by large models, extending beyond the bottlenecks of traditional NLP tools to explore how to effectively leverage \llm{} for security research.

\vspace {1pt}\noindent$\bullet$\space\textbf{\textit{New approach.}}
We introduce an automated tool that leverages \llm{} to extract API usage patterns from official documentation.
We are inspired by the ReAct framework to decompose the RM-API constraint retrieval into two parts with additional expertise: 1) allocation APIs identification which adopts a two-dimensional prompting method for cross-validation and an inconsistency-checking approach between the answer and the reasoning process; 2) releasing APIs pairing which adopts pre-processed function declaration for diminishing hallucinations.
Compared with the state-of-the-art API misuse detectors, \chatd{} is able to discover at least 80.85\% more RM-API pairs beyond expert knowledge and achieved an accuracy of 98.21\%. When applied to six popular libraries and the applications that use them, \chatd{} discovered 115 security issues and were ethically reported to the developers.
%说明已经ethically disclosure

\vspace {1pt}\noindent$\bullet$\space\textbf{\textit{Insightful Findings.}}
Through the analysis of the detection results of \chatd{}, we found that functions with lower usage frequency correspond to relatively incomplete documentation descriptions. Additionally, we observed that among the resource-allocation functions identified by \chatd{}, only 52.17\% of the functions had RM sentences with commonly used keywords. 
Although the inspiring results show that \llm{} has assisted in discovering more constraints beyond expertise, the small number of false positives introduced by this black-box model is currently challenging to explain which needs to be addressed in future research.

\section{Background and Preliminary Study}\label{sec:backandstudy}

\subsection{Background}\label{sec:back}

%\vspace{-5pt}
\noindent\textbf{Large Language Models.}
Large Language Models (LLMs), trained on vast corpora, exhibit proficiency across diverse natural language processing tasks, including common sense question-answering and semantic extraction~\cite{feng2023prompting}. The emergence of models like \chat{}~\cite{brown2020language} has brought heightened attention to both the security issues inherent in \llm{} and their potential applications in addressing security challenges.
Among these, \chat{} stands out as a state-of-the-art model designed for general-purpose tasks, initially based on the GPT-3.5 architecture and even more effective GPT-4~\cite{bubeck2023sparks}. Taking both cost and performance into consideration, we opted to employ GPT-3.5 for implementation.

%However, OpenAI later introduced the even more potent , surpassing GPT-3.5 in language understanding while also expanding its analytical prowess to encompass other data types, including code. However, faced with challenges in harnessing the GPT-4 model for large-scale automated calls, in this endeavour, we opted to employ GPT-3.5 for implementation.

%\vspace{-5pt}
\noindent\textbf{Reasoning with Large Language Models.}
Initially, recognizing the limitations of \llm{} in tackling complex problems, researchers introduced the concept of the \textit{Chain-of-Thought} (CoT) approach~\cite{wei2022chain}. This methodology guides \llm{} through a step-by-step thought process, empowering them to effectively address complicated problems and generate precise answers.
However, conventional CoT methods have exhibited considerably lower effectiveness in resolving textual reasoning tasks in comparison to math tasks. Consequently, researchers have introduced a novel framework known as ReAct framework~\cite{yao2023react}. By combining CoT with external knowledge, the \textit{``Think-Act-Observe''} framework is employed for task decomposition.
Inspired by these, our approach breaks down the complex task of recognizing RM-API constraints by supplying additional documentation content to the \llm{}. This augmentation enables efficient identification and precise matching of RM-APIs.

%\vspace{-5pt}
\noindent\textbf{CodeQL.}
CodeQL~\cite{CodeQL} is a static analysis tool for detecting security issues with QL queries written by developers or on users' own.
The analysis procedure consists of three steps: creating a CodeQL database, running QL queries against the database, and interpreting the analysis results.
CodeQL generates a database by extracting relational representations of source files, monitoring build processes for compiled languages, and resolving dependencies for interpreted ones. Each supported language has its extractor to maintain accuracy, and databases are created one language at a time before importing data into a CodeQL database directory.
Once a CodeQL database is created, developers execute QL language queries against it. 
%These queries, either from the CodeQL repository or custom ones, can be run using the CodeQL for VS Code extension or the CodeQL CLI.
In the final step, results from query execution are interpreted to highlight potential issues in the source code, based on metadata properties within the queries. 
%The interpreted results are then displayed for code review and triaging, with CodeQL for Visual Studio Code showing them automatically in the source code, while the CodeQL CLI offers various output formats for compatibility with different tools.

%%%%%%%%Framework%%%%%%%%%%
\begin{figure*}[t]
    \centering
    \includegraphics[width=1.02\textwidth]{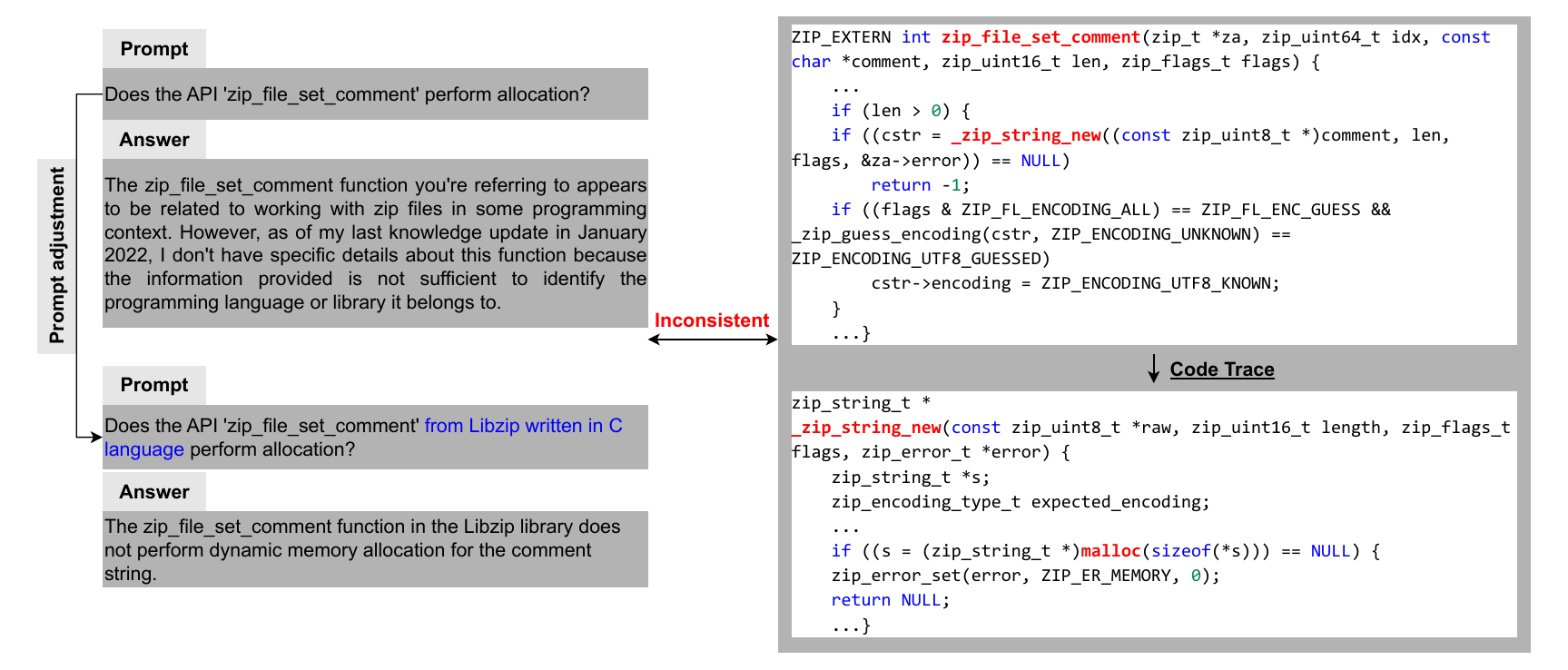}
    \caption{The limitation for direct-use of \llm{} for RM-API Identification}  
    \label{fig:no4api-iden}
    %\vspace{5pt}
\end{figure*}
%%%%%%%%Framework%%%%%%%%%%

%%%%exp-api-pairing%%%%%%
\begin{figure}[t]
    \centering
    \includegraphics[width=0.48\textwidth]{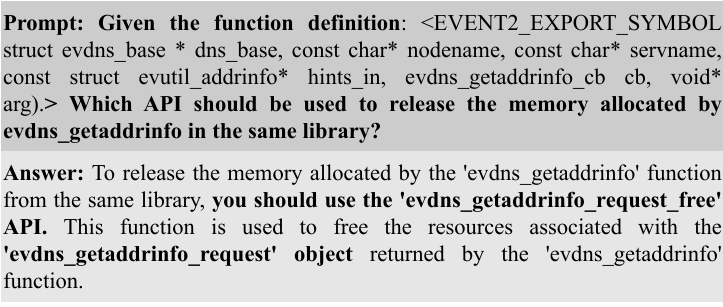}
    \caption{The limitation for direct-use of \llm{} for RM-API pairing}
    \label{fig:noGPT4pairing}
    \vspace{-5pt}
\end{figure}
%%%%%%%%%%%%%%%%%%%%%%%%%

\subsection{Preliminary Study}\label{sec:pre-study}
%把直接使用LLM的例子列出来，并解释相关challenge
%把方法中的与origional llm对比的部分提出来，单独写
In this section, we aim to tackle two tasks: allocation-API identification and RM-API pairing, directly leveraging large language models (LLMs). We explicitly outline the challenges and specific issues encountered when employing this approach.

%\vspace{-5pt}
\noindent\textbf{\ding{182} \llm{} fabricate answers without expertise.}
We initially prompt \llm{} for identifying RM APIs, as depicted in Figure~\ref{fig:no4api-iden}. \llm{} responds that it finds it difficult to provide an answer due to insufficient information. Upon refining the prompt, \llm{} indicates that the current API \texttt{zip\_file\_set\_comment} is not involved in allocation. However, upon inspecting the source code, it is revealed that \texttt{zip\_file\_set\_comment} indirectly invokes \texttt{malloc}, indicating its allocation semantics, contrary to the assertion made by \llm{}.
%Therefore, we assume the first challenge in directly utilizing LLMs for RM-API identification is \textbf{} 
The same phenomenon also exists in RM-API Pairing. 
We directly utilize \llm{} for releasing API retrieval as illustrated in Figure~\ref{fig:noGPT4pairing}. When providing the function declaration of \texttt{evdns\_getaddrinfo} to \llm{} and asking for the corresponding resource-releasing API, the \llm{} outputs the result \texttt{evdns\_getaddrinfo\_request\_free}. As shown in the figure, \llm{} not only gives the \textit{function description} of this API but also provides the object type that needs to be released (RM Object Type). This seemingly reasonable answer and explanation are, in fact, incorrect. The truth is that the function \texttt{evdns\_getaddrinfo\_request} does not exist and was fabricated by \llm{}. Such a phenomenon could also be one of the impacts introduced by \llm{}'s hallucinations. Due to the lack of expertise, \llm{} fabricates answers based on prior experience or stereotypes. For instance, \llm{} assumes that a resource-releasing API should contain the keyword \textit{free}, and the corresponding API should contain partially identical keywords such as \textit{evdns\_getaddrinfo}. 
Therefore, we propose to perform reasoning with \llm{} and provide them with external expertise to improve the accuracy of downstream tasks.

%\vspace{-5pt}
\noindent\textbf{\ding{183} \textbf{\llm{} generate incorrect answers with evidence.}}
Another challenge occurs in the reasoning process, for example, when providing the descriptions of \texttt{zip\_discard} \textit{``The zip\_discard() function closes the archive and frees the memory allocated for it''} to \llm{}, the result of the prompt \textit{Does this API perform allocation?} is ``YES'' which is contrary to the fact. 
%分析产生这一问题的原因
One possible reason for these cases is that hallucinations exist in \llm{}~\cite{wei2022emergent}. According to the recent research~\cite{agrawal2023knowledge}, the reasons for this phenomenon are typically a lack of context, leading to an inability to understand the questions or a lack of expertise resulting in inaccurate answers.
Therefore, we propose an inconsistency-checking approach between the \llm{}' output and the reasoning process to achieve better performance in RM-API Identification.

\section{Design and Implementation} \label{sec:design}

%overview
%%%%%%%%Framework%%%%%%%%%%
\begin{figure*}[t]
    \centering
    \includegraphics[width=0.8\textwidth]{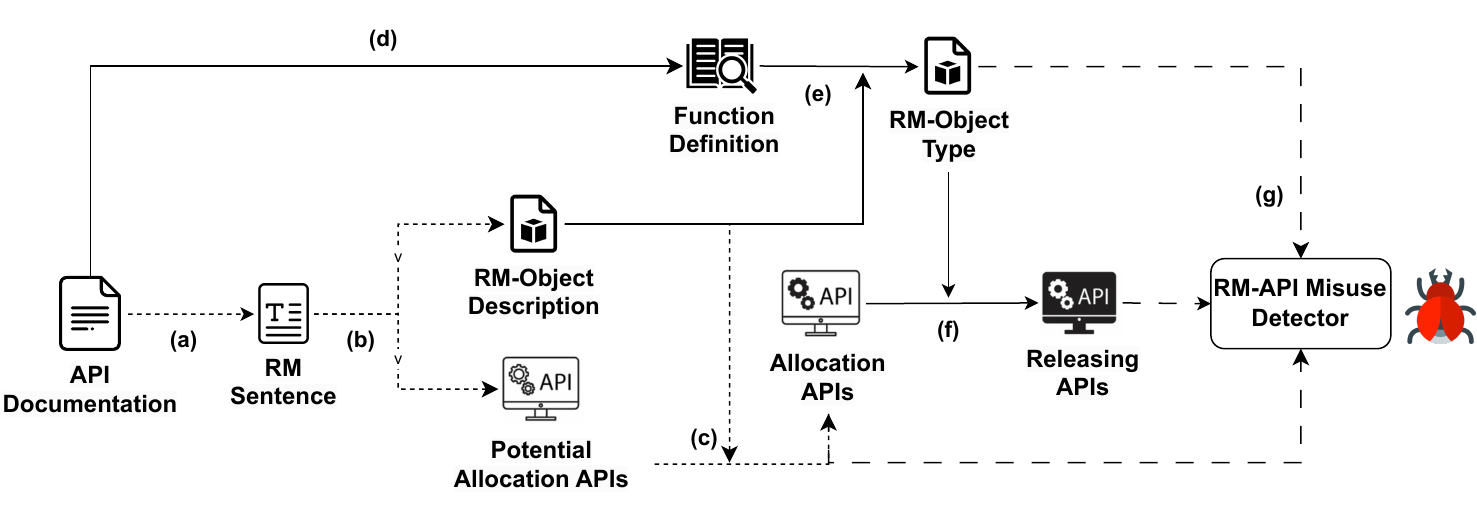}
    \caption{Architecture of \chatd{}} 
    \label{fig:arc-llm-pair}
\end{figure*}
%%%%%%%%Framework%%%%%%%%%%

%%%%%%%%%%%%%%%%%%%%%%%%%%
%RM语句示例
\begin{table*}[t]
\centering
\small
\setlength{\belowcaptionskip}{10pt}
\caption{Examples of RM Sentences}
\label{tab:rm-sents-cla}
\begin{tabular}{>{\rule{0pt}{2.3ex}}m{1.1cm}|m{9.0cm}|m{3.2cm}|m{3.2cm}}
\hline
 &\textbf{RM Sentences} & \textbf{RM API}  & \textbf{RM object Description}  \\
\hline
\textbf{Type-I} & ``The functions \textbf{zip\_source\_filep() and zip\_source\_filep\_create()} create \textbf{a zip source} from a file stream.''& zip\_source\_filep() and zip\_source\_filep\_create() & a zip source\\
\hline
\textbf{Type-II} & ``\textbf{a newly allocated struct event} that must later be freed with event\_free() or NULL if an error occurred.''& - & struct event\\
\hline
\end{tabular}
\vspace{-10pt}
\end{table*}
%%%%%%%%%%%%%%%%%%%%%%%%%%%%%%%%%%%%%%%

\subsection{Overview}
In this section, we provide a detailed explanation of our approach, \chatd{}, which can analyse official documentation with \llm{} to extract RM-API constraints and achieve RM-API misuse detection. \revise{We also provide a description of the overall framework and then illustrate each component with detailed examples.}

%\vspace{-5pt}
\noindent\textbf{Framework.}
This work aims to explore the capabilities of \llm{} in RM-API constraints retrieval and RM-API misuse detection.
As illustrated in Figure~\ref{fig:arc-llm-pair}, \chatd{} consists of three main components: utilizing \llm{} to analyze documentation for resource-allocation APIs identification (Figure~\ref{fig:arc-llm-pair} (a-c)), recognizing corresponding resource-releasing APIs with \llm{} (Figure~\ref{fig:arc-llm-pair} (d-f)), and performing RM-API misuse detection using CodeQL based on the identified RM-API pairs (Figure~\ref{fig:arc-llm-pair} (g)). \chatd{} adopts \chat{} (GPT-3.5-turbo model~\cite{gpt-3.5}), one of the state-of-the-art \llm{} due to its superior performance and cost-effectiveness (detailed discussion are shown in Section~\ref{sec:discuss}), for the whole implementation.
The whole process of \llm{} utilization is fully automated and human assistance is only needed in detection code generation when using CodeQL for misuse detection, which is a one-time effort.  
More specifically, inspired by ReAct framework, \chatd{} decomposes the allocation-API identification into two parts. First, it identifies the RM sentences from the official documentation describing the current API's functionality and possessing resource-allocation semantics. Then, adopting a two-dimensional prompting method for cross-validation, \chatd{} prompts to provide potential resource-allocation APIs and descriptions of the allocated objects (RM objects). Through the inconsistency-checking between the \llm{}' output and the reasoning process, APIs with allocation semantics are obtained.
Next, by utilizing the pre-processed function declarations, \chatd{} can extract the corresponding RM object types and output the corresponding resource-releasing APIs with diminished hallucinations.
Finally, the identified RM-API pairs are written into QL queries, and RM-API misuse detection is performed using CodeQL~\cite{CodeQL} on open-source software.

\ignore{
%%%%%%%%%%%%%%%%%%
%example举例
\begin{figure*}
    \centering
    \includegraphics[width=0.35\textwidth]{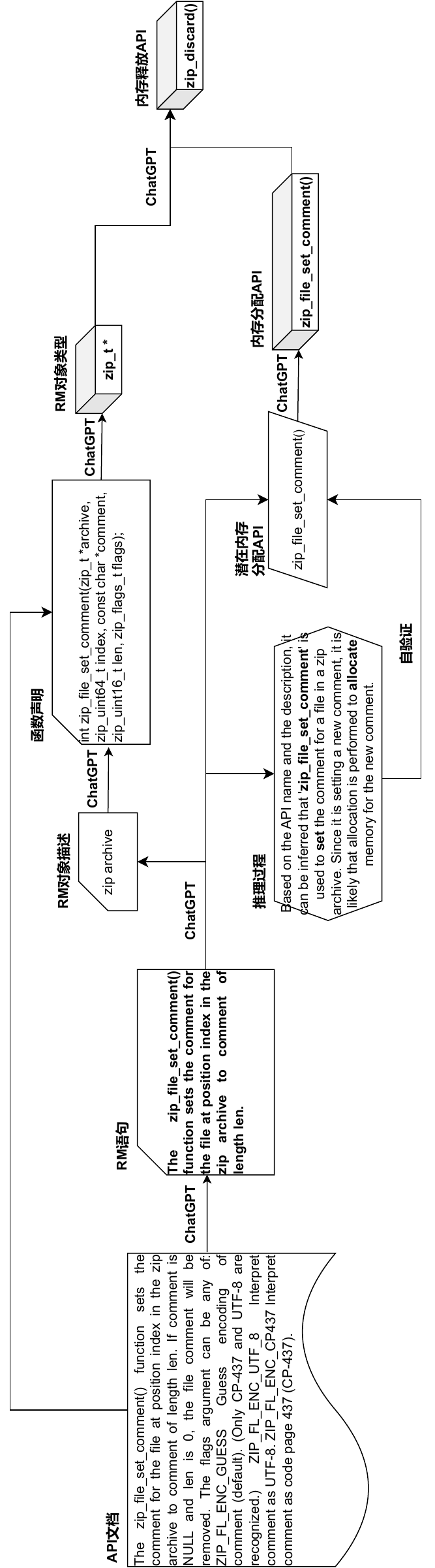}
    \caption{Example of how \chatd{} works.}
    \label{fig:exp-all}
\end{figure*}
%%%%%%%%Framework%%%%%%%%%%
}

\ignore{
\paragraph{Example.}
Figure~\ref{fig:exp-all} illustrates how \chatd{} achieves RM-API pairs identification.
\todo{add more details.}
}

\subsection{RM-API Identification}
\label{sec:gpt_api_iden}

\vspace{-5pt}
\paragraph{API Semantics Confirmation.}
\revise{We \yy{provide} a detailed definition for the RM-APIs.} Specifically, we consider the APIs that call the \texttt{malloc} function to have allocation semantics; and the APIs that call the \texttt{free} function to have releasing semantics. 
\revise{We manually check each API's source code and documentation according to the procedure proposed in Section~\ref{sec:explore}. 
To further confirm the allocation semantics of the API, we obtain the allocated object of the current API through data-flow analysis and control-flow analysis. If the allocated object needs an extra releasing operation outside the API, we assume the current API has allocation semantics.}
%具体来说，通过人工交叉验证，我们获取了核心调用了malloc/free的API，以及API内部被allocate以及被release的对象。在数据依赖分析和控制依赖分析的辅助下，获取当前API被allocated的对象，并确定该对象需要在API外被额外释放，从而认定当前API是具有malloc语义的。
\ignore{
We use a brief counting method for quick double-verification assistance during the construction of GroundTruth.
Also for the APIs that call both \texttt{malloc} and \texttt{free} functions, we consider the APIs to have allocation semantics when the number of \texttt{malloc} calls exceeds the number of \texttt{free} calls; we consider the APIs to have releasing semantics when the number of \texttt{free} calls exceeds the number of \texttt{malloc} calls.}
The semantics of the specific API is confirmed by the latest official documentation and source code. Besides, the semantics of a function is fixed from the outset and in most cases does not change due to iteration. 
In other words, when a function is identified to have allocation semantics, its functionality will not change due to the iteration of function versions, that is, it will not switch from having allocation semantics to having releasing semantics.

\vspace{-5pt}
\paragraph{The Definitions.} 
Through analysis of a large amount of API documentation, we observe that APIs with \textbf{RM semantics} (such as, allocate/release) usually have explicit or implicit descriptions within the documentation. 
We term the sentences that have the semantics of such APIs as \textbf{RM sentences}. \revise{More specifically, \yy{if an API} calls malloc/free in the source code and its description \yy{includes} the keywords that are semantically similar to malloc/free, \yy{we treat this sentence} as RM sentences, the keywords that represent malloc semantics fall in ``create'', ``allocate'', ``construct'' and ``initialize'' etc. We then construct the GroundTruth of 244 sentences through manual inspection from three libraries including FFmpeg, Libdbus and Libpcap.}
\revise{The statistics show that 18.4\% of RM sentences contain function names and 81.6\% of them contain the allocated/released objects. Based on this, we classify the RM sentences} into two types, as illustrated in Table~\ref{tab:rm-sents-cla}.
The first type of RM sentence is extracted from the description of the function \texttt{zip\_source\_filep}. This sentence explicitly states that the function containing memory allocation functionality and the memory allocated is stored in ``a zip source'', which we term as the \textbf{RM object description}.
%This sentence unmistakably indicates that the function \texttt{zip\_source\_filep} exhibits allocation semantics.
The second type of RM sentence is extracted from the return value description of the function \texttt{event\_new}. This sentence implies the memory allocation semantics of the function by stating that the return value \textit{struct event}, which can be referred to as \textbf{RM object}, represents the allocated memory and requires releasing using another function. 
Although extracting the RM-API name directly from this sentence is challenging, we can infer the RM object described in this sentence by understanding its semantic context.

From the preliminary study in Section~\ref{sec:pre-study}, directly using \llm{} for RM-API identification results in many incorrect answers. We also explored adding more information to identify API's semantics with source code or documentation(shown in Section~\ref{sec:RQ3}), but neither achieved satisfactory results in the current task. 
Therefore, inspired by the ReAct framework~\cite{yao2023react} optimizing on \textit{Chain-of-Thought}~\cite{wei2022chain}, we propose to decompose the complex task (RM-API constraint retrieval) into RM sentence extraction and allocation APIs identification, together with additional information retrieved from API documentation.

\begin{figure}[t]
    \centering
    \includegraphics[width=0.46\textwidth]{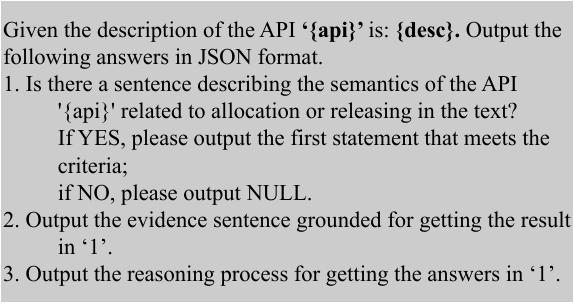}
    \caption{The prompt for extracting RM Sentences}
    \label{fig:rm-sents-extract}
    \vspace{5pt}
\end{figure}

\vspace{-5pt}
\paragraph{RM Sentence Extraction.}
According to the definitions above, RM sentence describes the functionality of the RM APIs and \chatd{} uses the prompt in Figure~\ref{fig:rm-sents-extract} to extract them from API documentation descriptions. Note that all the content such as the \textit{api} and \textit{desc} representing description, highlighted in brackets are automatically crawled from official documentation. The rest context in the prompt is the template which is a one-time effort.
The procedure is illustrated as follows.
First, \chatd{} prompts the \llm{} to provide the first sentence in the documentation describing the current API's functionality and check if it possesses RM semantics. When multiple API descriptions are present on the same documentation page, the sentence outlining a particular API's functionality typically resides close to that API. In cases where documentation exclusively details a single API and contains multiple sentences with RM semantics, to prevent redundancy, we exclusively select the first sentence that delineates the API's functionality and exhibits RM semantics as the RM sentence.
Secondly, owing to the potential issue of \textit{hallucination} in the output of \llm{}~\cite{maynez2020faithfulness}, it is imperative to verify whether the presently identified RM sentence originates from the documentation itself and is not fabricated. Furthermore, confirming the accuracy of this response is essential. Consequently, \chatd{} prompts \llm{} to provide the reasoning process and the source-based evidence of RM sentence identification to support the answer.

%%%%exp-api-pairing%%%%%%
%%%%%%%%%%%%%%%%%%%%%%%%%

%%%%exp-api-pairing%%%%%%
\begin{figure}[t]
    \centering
    \includegraphics[width=0.47\textwidth]{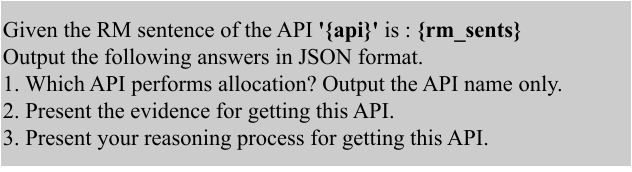}
    \caption{The prompt (a) for identifying RM-APIs}
    \label{fig:rm-api-iden-a}
    %\vspace{-10pt}
\end{figure}
%%%%%%%%%%%%%%%%%%%%%%%%%

%%%%exp-api-pairing%%%%%%
\begin{figure}[t]
    \centering
    \includegraphics[width=0.47\textwidth]{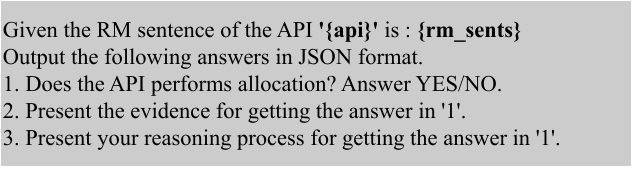}
    \caption{The prompt(b) for identifying RM-APIs}
    \label{fig:rm-api-iden-b}
\end{figure}
%%%%%%%%%%%%%%%%%%%%%%%%%

%%%%exp-api-pairing%%%%%%
\begin{figure}[t]
    \centering
    \includegraphics[width=0.47\textwidth]{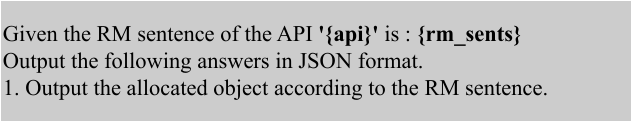}
    \caption{The prompt for identifying RM-objects}
    \label{fig:rm-obj-iden}
    %\vspace{-10pt}
\end{figure}
%%%%%%%%%%%%%%%%%%%%%%%%%

\vspace{-5pt}
\paragraph{Allocation APIs Identification.}
To perform allocation APIs identification, we tried to directly apply the extracted RM sentences to the prompt. However, due to the inherent uncertainty introduced by \llm{}, accurately validating the truthfulness and accuracy of current answers using a single prompt poses challenges. 
We observed the RM sentences can be classified into two categories: those containing both RM-API names and RM objects, and those containing RM objects only.
For sentences containing RM objects only, RM APIs identification can be achieved by assessing the presence of RM objects and verifying the accuracy of \llm{}'s response.
On the other hand, for those sentences containing both RM-API names and RM objects, RM APIs identification necessitates additional validation for authenticity and accuracy, building upon the previous step.
And also, the previous studies show that \llm{} are effective in three types of prompting methods, which are \textit{in-context prompting}: extracting answers from the given context and the answers can be extracted from the input context; \textit{open prompting}: generating answers based on the information from the given context and the answers cannot be extracted from the input context; and \textit{indirect prompting}: generating answers based on the indirect questions, both the prompts and the answers pose indirect relations to the final answers.
Therefore, \chatd{} follows the above three types of prompting and adopts the observation to design the prompts in Figure~\ref{fig:rm-api-iden-a} and Figure~\ref{fig:rm-api-iden-b} to prompt \llm{} for allocation APIs identification based on RM sentences, and utilizes the prompt in Figure~\ref{fig:rm-obj-iden} to obtain RM-objects description.

%%%%exp-api-pairing%%%%%%
\begin{figure}[t]
    \centering
    \includegraphics[width=0.47\textwidth]{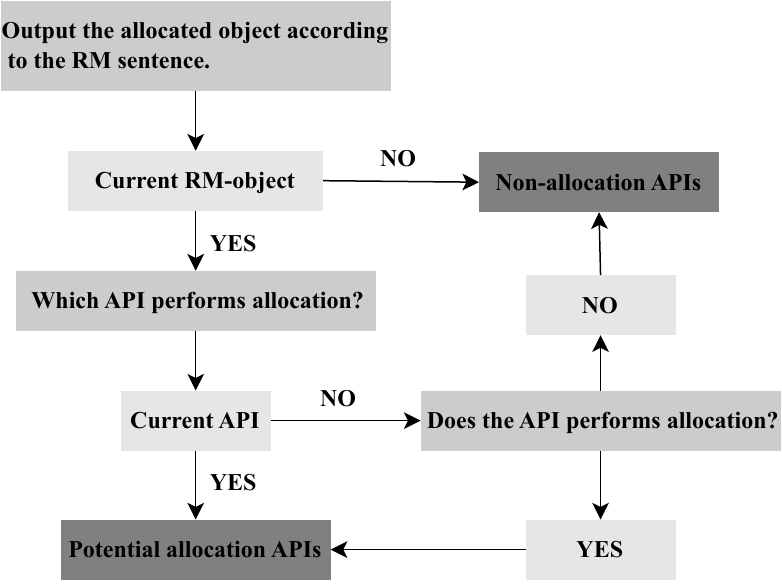}
    \caption{The rules for Cross-Validation}
    \label{fig:validate-rules}
    %\vspace{10pt}
\end{figure}
%%%%%%%%%%%%%%%%%%%%%%%%%

Specifically, the prompt in Figure~\ref{fig:rm-api-iden-a} is designed to inquire about the API with resource allocation functionality based on the reasoning derived from the current RM sentence. Additionally, the prompt in Figure~\ref{fig:rm-api-iden-b} is designed to ascertain whether the current API possesses resource allocation semantics. And the prompt in Figure~\ref{fig:rm-obj-iden} is intended to acquire the RM-object.
To further confirm the allocation APIs, we design rules (details are shown in Figure~\ref{fig:validate-rules}) to cross-validate the three answers. The necessary condition is to ensure the extracted RM objects exist in the documentation; if the response to the prompt in Figure~\ref{fig:rm-api-iden-a} is the \texttt{current API} or the response to the prompt in Figure~\ref{fig:rm-api-iden-b} is \texttt{YES}, then \chatd{} collects the potential allocations APIs through the cross-validation.

However, the current method leads to many false positives, primarily due to \chat{} being a black-box model, making it challenging to be guided.
Besides, we observe typical self-inconsistencies between the \llm{}'s output and the reasoning processes. For example, \llm{} indicate that \texttt{zip\_close} has memory allocation semantics. But the reasoning process for \texttt{zip\_close} is as follows: \textit{``The sentence mentions that the zip\_close function \textbf{frees} the archive, indicating that its purpose is to \textbf{deallocate} resources rather than allocate them. Therefore, it can be inferred that the function zip\_close \textbf{does not perform allocation}.''}
The reasoning process confirms the identified RM sentence is correct, but \llm{} has limitations in understanding this sentence. Even when the RM sentence clearly states that the current function's operation is memory deallocation, the model still interprets the function's semantics as the opposite. Therefore, we utilize an off-the-shelf NLP tool, Semantic Role Labeling~\cite{parsing2009speech}, to extract the action verbs corresponding to the current function within the reasoning process, such as \textit{set}, \textit{free} and others. By comparing the semantic similarity between these action verbs and \textit{release}, \textit{allocate} or \textit{Neutral}, \chatd{} can further determine the semantic tendency of the current function. Then we acquire the resource-allocation APIs.
%Through the analysis of API documentation of three popular libraries, 
%\chatd{} identifies 88 resource-allocation APIs, achieving an accuracy of 93.62\% and a recall of 95.65\% (details are shown in Section~\ref{sec:evaluation}).
%\chatd{} identifies 168 resource-allocation APIs, achieving an accuracy of 92.81\% and a recall of 92.06\% (details are shown in Section~\ref{sec:evaluation})

%%%%exp-api-pairing%%%%%%
\begin{figure}[t]
    \centering
    \includegraphics[width=0.5\textwidth]{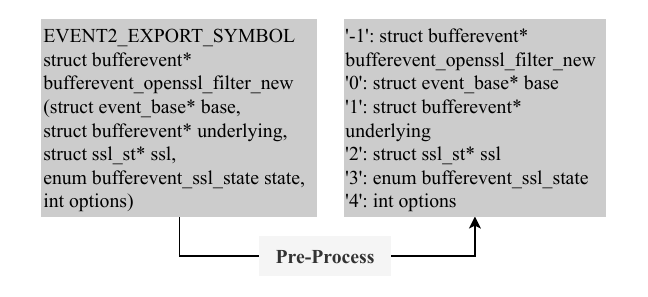}
    \caption{The pre-process procedure of Function Definition}
    \label{fig:func-pre}
    %\vspace{-5pt}
\end{figure}
%%%%%%%%%%%%%%%%%%%%%%%%%

%%%%exp-api-pairing%%%%%%
\begin{figure}[t]
    \centering
    \includegraphics[width=0.47\textwidth]{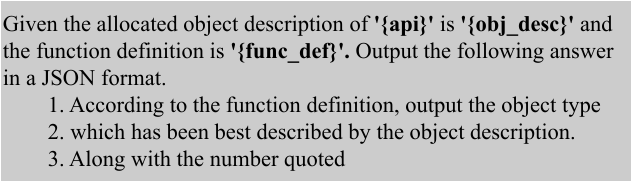}
    \caption{The prompt for RM-object Identification}
    \label{fig:rm-obj-type-iden}
    %\vspace{5pt}
\end{figure}
%%%%%%%%%%%%%%%%%%%%%%%%%

%%%%free-api pairing%%%%%%
\begin{figure}[t]
    \centering
    \includegraphics[width=0.47\textwidth]{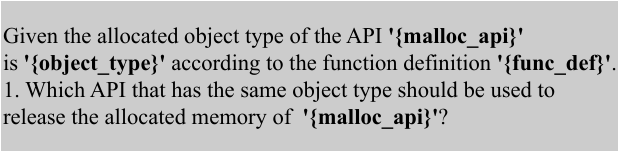}
    \caption{The prompt for Releasing API Identification}
    \label{fig:f-api-iden}
    %\vspace{-5pt}
\end{figure}
%%%%%%%%%%%%%%%%%%%%%%%%%

\subsection{RM-API Pairing}
\begin{figure}[t]
    \centering
    \includegraphics[width=0.47\textwidth]{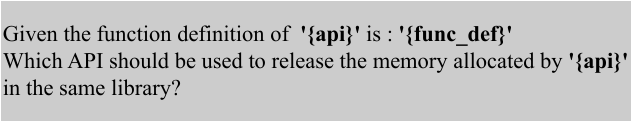}
    \caption{The prompt for initial releasing API Identification}
    \label{fig:m-api-pair}
    %\vspace{-10pt}
\end{figure}

According to the preliminary study in Section~\ref{sec:pre-study}, we attempted to directly apply the obtained allocation APIs along with the function definition for corresponding releasing APIs identification by using the prompt in Figure~\ref{fig:m-api-pair}. However, due to the inherent \textit{hallucinations} in \llm{}, a great number of fabricated answers were generated. 
\llm{} might fabricate the answers due to the inability to solve complex tasks and insufficient information for simple task completion. 
The common knowledge says that the RM-API pairs share the same RM-object type. That is, the released object of the releasing API is typically the allocated object of the allocation API, the naming of these two APIs may differ, but the type of RM object remains consistent. 
Based on this knowledge, \chatd{} adopts the inspiration from ReAct framework to decompose the releasing APIs identification task into RM-object Type Extraction and releasing API Identification.

%Based on this observation, we attempt to use RM object descriptions to locate the type of RM object in the function declaration of resource allocation APIs. We then provide additional prompts to \chat{} for the RM object type to obtain the corresponding resource deallocation API.

%%%%exp-api-pairing%%%%%%
%%%%%%%%%%%%%%%%%%%%%%%%%

\noindent\textbf{RM-object Type Extraction.}
\label{para: obj_iden}
To achieve RM-object Type Extraction, we first attempted to prompt \llm{} with RM object descriptions retrieved from Section~\ref{sec:gpt_api_iden} and function declarations which were automatically retrieved from API documentation. However, this approach did not effectively enable \llm{} to identify the RM object's type. For example, when trying the prompt in Figure~\ref{fig:m-api-pair} for releasing API identification, the output of \llm{} tended to confuse the function names and types, which may result from insufficient information in the training set of the model. 
As a solution, \chatd{} pre-processes the function declarations by separating the types and names of parameters/return values, the details are illustrated in Figure~\ref{fig:func-pre}. Specifically, for \texttt{bufferevent\_openssl\_filter\_new} with its complete declaration, \chatd{} pre-processes it by splitting the types and names of parameters and return values along with numeric markers. The pre-processed declaration is shown in the figure. 
Further, \chatd{} uses the prompt in Figure~\ref{fig:rm-obj-type-iden} to extract the corresponding RM object type.
Experiments have shown that breaking down lengthy function declarations into parts can enhance \chat{}'s accuracy in matching RM object types based on the function declaration.

\noindent\textbf{Releasing API Pairing.}
\label{para: free_api_iden}
Based on the extracted RM object type, \chatd{} further acquires the corresponding resource-releasing APIs by providing \llm{} with the identified allocation API and the RM-object type with the utilization of the prompt in Figure~\ref{fig:f-api-iden}. In this step, we did not design a complex solution. The reason is that the \textit{hallucination} issues observed in the initial design of \llm{} on the resource-allocation APIs Identification gradually disappeared as we continuously provided valid expert knowledge and complex task decomposition solutions.
Experiments in Section~\ref{sec:evaluation} have shown that \chatd{} can accurately identify the correct RM-API pairs. This result indicates that \textit{hallucination} can be further resolved through continuous interaction with expertise and the decomposition inspired from \textit{Chain-of-Thought}.

%when \chatd{} can accurately provide RM object types and resource-allocation APIs, \llm{} can achieve an accuracy rate of up to 96.59\% in providing the correct resource-releasing APIs. 

\subsection{RM-API misuse detection}
Based on the previously identified RM-API pairs, we have obtained resource-allocation APIs, resource-releasing APIs, and the shared RM object types. In this section, we will provide a detailed explanation of how to perform RM-API misuse detection using CodeQL~\cite{CodeQL}. For each RM-API pair, \chatd{} applies CodeQL to detect three types of security bugs: memory leaks, use-after-free, and double-free.
Applying CodeQL to security bug detection includes three steps (Section~\ref{sec:back}).
First, we create the database for each application integrating one of the six libraries (details are shown in Table~\ref{tab:lib-info} in Appendix). Then we generate corresponding QL queries based on the detected RM-API pairs (each pair corresponds to one piece of constraint) to detect RM-API misuse. 
\revise{Note that the queries are manually constructed based on the basic template provided by CodeQL \yy{and the construction procedure} is a one-time effort \yy{which} requires no need to modify if applied to detect the same bug type on new software.}
We then interpret the results, confirm the misuse and ethically report to developers.
For memory leak bugs, the tool locates a given allocation API and the allocated object. 
Based on the data-flow analysis and control-flow analysis, the tool checks every path within the code block to see if there is a corresponding releasing API to free the allocated object.
If there are paths that do not contain the corresponding releasing API, then a memory leak occurs.
For double-free bugs, the tool checks if 
one path contains two releasing APIs on the same object and there are no other allocation APIs between them which results in double-free bugs.
For use-after-free bugs, the tool checks whether there is an operation to dereference the freed object after the releasing API, if so, it is a use-after-free bug.

Take the memory-leak detection strategy of \texttt{evconnlistener\_new\_bind} as an example and the QL code is listed in Listing~\ref{lst:memleak-new-bind}.
CodeQL first calls the function \texttt{isSourceFC} to locate the control flow node by the function name (\texttt{evconnlistener\_new\_bind}), indicating the current block contains the function.
Then the tool calls the function \texttt{isLocalSource} to verify the RM object is a local variable.
Finally, CodeQL calls the function \texttt{getLeakBlock} which internally calls the function \texttt{getSourceExpr} to locate the node corresponding to the specified parameter of the function.
The method detects no post path of the function \texttt{evconnlistener\_new\_bind} contains the corresponding releasing function which indicates there exists a memory leak within the basic block.
CodeQL supports modifying the RM-API names and the corresponding indexes of RM objects to automatically generate corresponding detection code which helps in large-scale bug detection.

%The following code provides the detection of memory leaks for \texttt{evconnlistener\_new\_bind}.
%%%%%%%%%%%%%%%%%-----CodeQL-----%%%%%%%%%%%%%%%%%%%%%%%%%
\begin{lstlisting}[language=c,frame=trBL,caption= {QL code for detecting the memory leak of function evconnlistener\_new\_bind },abovecaptionskip=10pt,belowcaptionskip=0pt,captionpos=b,label={lst:memleak-new-bind}, keywordstyle=\bfseries\color{green!40!black},
commentstyle=\itshape\color{purple!40!black},
identifierstyle=\color{blue},
stringstyle=\color{orange},
basicstyle=\footnotesize\ttfamily,
breaklines=true,
numbers=left,
numbersep=-5pt,
stepnumber=1]
  predicate isSourceFC(FunctionCall fc)
  {
  // malloc function: evconnlistener\_new\_bind
  fc.getTarget().hasName("evconnlistener_new_bind")
  }
  
  Expr getSourceExpr(FunctionCall fc)
  {
  //parameter 1:
  result = fc.getArgument(0)
  //return value:
  //result = fc 
  }

  import cpp
  from BasicBlock bb, FunctionCall malloc
  where 
    // locate the malloc function call by allocation API name
    isSourceFC(malloc)
    //Make sure that the malloc function operates on a local variable
    and isLocalSource(malloc)
    //There is a path in the current malloc function postpath that does not have a releasing operation
    and bb = getLeakBlock(malloc)
  select malloc, malloc.getLocation().toString()
\end{lstlisting}
\vspace{-5pt}
%%%%%%%%%%%%%%%%%%%%%%%%%%%%%%%%%%%%%%%%%%%%%%%%%%
%\input{Sections/implementation.tex}
\section{Evaluation}\label{sec:evaluation}
%overview

\subsection{RQ1: Effectiveness}
%\todo{add intro descriptions}
In this section, we implement \chatd{} on open-source libraries and the applications integrated on them to further evaluate \chatd{}'s performance.  

\vspace{-5pt}
\paragraph{End-to-end Effectiveness.}\label{subsuc:eval}
We implement \chatd{} on six popular libraries documentation: FFmpeg~\cite{Ffmpeg}, Libevent~\cite{libevent},
Libexpat~\cite{libexpat},
 Libpcap~\cite{Libpcap}, Libzip~\cite{Libzip} and OpenLdap~\cite{Openldap}  (details are shown in Table~\ref{tab:lib-info} in the Appendix), and the Top 10 applications integrating on them.
In total, \chatd{} discovered 165 RM-API pairs \revise{containing 165 allocations APIs and 61 releasing APIs, for some of the allocation APIs may have multiple releasing APIs to release different parameters and some of them may have the same releasing API for multiple allocation APIs.} \revise{We followed the ethical bug disclosure policy of software and reported all of the }115 security bugs, \revise{64 of which were confirmed by the developers} \revise{with a precision rate of 90.2\%. And the disclosure procedure has no ethical issues.} The details are shown in Table~\ref{tab:diclosed_misuses} in the Appendix.
Figure~\ref{fig:codeql-fp-1} illustrates a false positive case in CodeQL's detection when inspecting the use of \texttt{zip\_source\_buffer}. \chatd{} detected the correct RM-API pair: the resource-allocation API \texttt{zip\_source\_buffer} and the resource-releasing API \texttt{zip\_source\_free}. CodeQL's detection revealed that no subsequent call exists to \texttt{zip\_source\_free} to release the allocated resource after calling \texttt{zip\_source\_buffer}. However, as depicted in the figure, the parameter \textit{res} is generated by the functions \texttt{zip\_file\_replace} or \texttt{zip\_file\_add}, and the documentation explicitly states, \textit{``NOTE: zip\_source\_free(3) should not be called on a source after it was used successfully in a zip\_file\_add or zip\_file\_replace call.''} Consequently, in the condition \textit{``res == -1''}, which corresponds to a call failure, \texttt{zip\_source\_free} is invoked to release resource. Due to CodeQL's inability to handle vulnerability detection when multiple functions collaborate, it resulted in a false memory leak detection.

%%%%%%%%%%%%%%%%%%%%%%%%%
%codeql fp
\begin{figure}[t]
    \centering
    \includegraphics[width=0.44\textwidth]{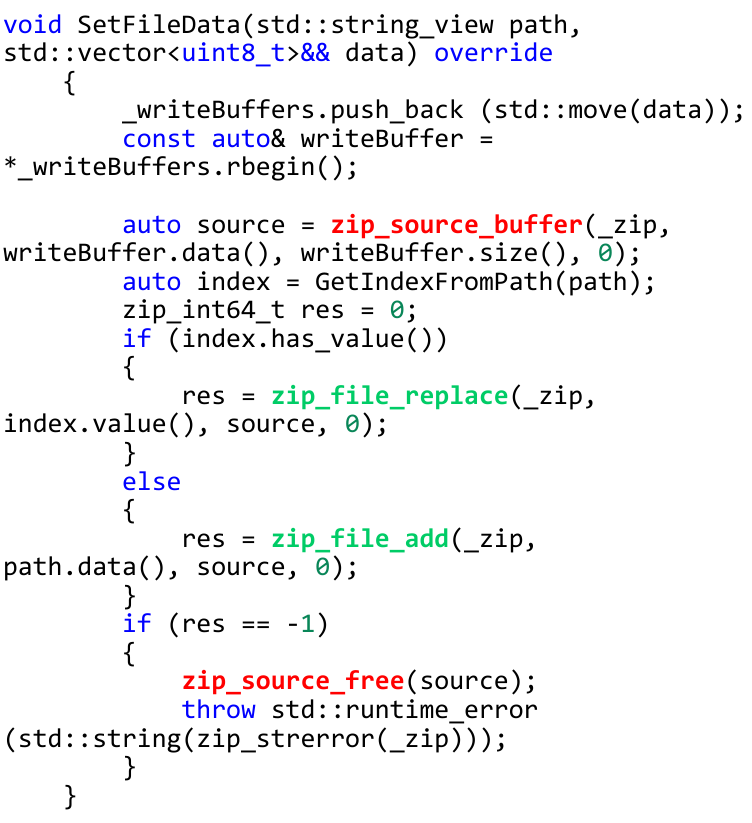}
    \caption{A false positive introduced by CodeQL}
    \label{fig:codeql-fp-1}
    \vspace{-5pt}
\end{figure}
%%%%%%%%%%%%%%%%%%%%%%%%%

\vspace{-5pt}
\paragraph{Effectiveness of RM-API Identification.}
%\chatd{} detected 88 resource allocation APIs across the three software libraries with an accuracy of 93.62\% and a recall of 95.65\%. Four false negatives
\revise{\chatd{} detectes 168 resource allocation APIs across the six software libraries with an accuracy of 92.81\% and a recall of 92.06\%. 14 false negatives} generated during the identification process were primarily due to the prior misidentification of RM sentences. This could be attributed to \llm{}' difficulty in locating key sentences in less obvious locations. For example, the documentation description for the function \texttt{zip\_open\_from\_source} is illustrated in Figure~\ref{fig:zip-open-doc} (details are shown in Appendix).
In this description, the sentences related to the \texttt{zip\_open\_from\_source} function do not appear at the beginning of the paragraph or are not prominently highlighted. This leads \llm{} to provide the incorrect RM sentence \textit{``The semantics of the API zip\_open\_from\_source related to allocation or releasing is not described in the given description''}. However, when we prompt \llm{} to determine the function's semantics based on the sentence \textit{``The zip\_open\_from\_source() function opens a zip archive encapsulated by the zip\_source zs using the provided flags. In case of error, the zip\_error ze is filled in''}, \llm{} is able to correctly identify that the current function has allocation semantics. Therefore, we attribute the occurrence of these false negatives to the inherent limitations of \llm{}' capabilities.

\revise{Simultaneously, \chatd{} also introduced 13 false positives.} Through the analysis of these cases, we found that the content in the documentation can also mislead \llm{} to make incorrect judgments. For instance, in the documentation description of the function \texttt{zip\_source\_begin\_write\_cloning} (shown in Figure~\ref{fig:zip-begin-doc} in the Appendix), the phrase \textit{``Usually this involves creating temporary files or allocating buffers''} explicitly indicates that the current function involves resource allocation operations. However, upon inspecting the source code, it was found that the function did not call any other functions involving \texttt{malloc}. Therefore, the inconsistency between the documentation description and the source code can also lead to false positives in \chatd{}, making these cases of misleading documentation errors.

%%%%exp-api-pairing%%%%%%
\begin{figure}[t]
    \centering
    \includegraphics[width=0.5\textwidth]{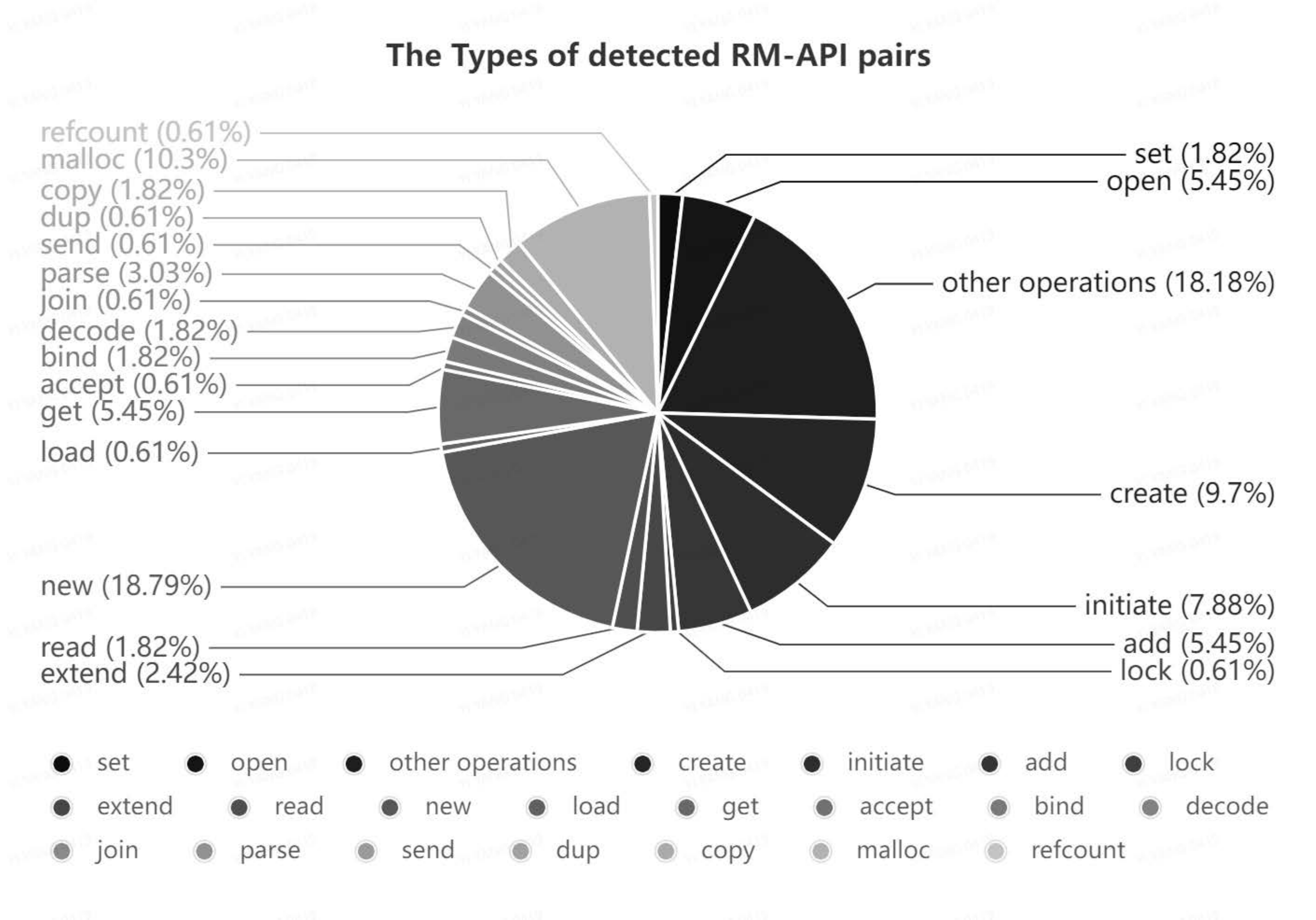}
    \caption{The types of detected RM-API pairs}
    \label{fig:type-pairs}
    %\vspace{5pt}
\end{figure}
%%%%%%%%%%%%%%%%%%%%%%%%%

\noindent\textbf{Effectiveness of RM-API Pairing.}
%\chatd{} detects 85 correct RM-API pairs in three software libraries (details are shown in Table~\ref{tab:effect-3tools}), achieving an accuracy of 96.59\%. The three false positives 
\revise{\chatd{} detects 165 correct RM-API pairs in six software libraries achieving an accuracy of 98.21\%.} The three false positives encountered during the matching process were primarily due to the lack of common sense in the \llm{} and the \textit{hallucinations} of the model.
For example, in the function \texttt{evhttp\_uri\_join}, which allocates memory and stores it in the return value, with the corresponding RM object type \textit{char*}. However, even though \chatd{} accurately identified the RM object type, for \llm{}, it was difficult to find the corresponding resource release function based on this general type. Therefore, the answer given by \chatd{} was \texttt{evhttp\_uri\_free}, which has a similar naming pattern to the function \texttt{evhttp\_uri\_join}, which however, was fabricated by the model.
In the case of the function \texttt{evwatch\_check\_new}, which allocates memory and stores it in \textit{struct evwatch*}, \llm{} provided a fabricated answer, \texttt{evwatch\_check\_free}, based on the correct RM object type. While this function had a similar naming pattern, it did not actually exist.
Although such cases are not common in the results of the current method, these random false positives also result in additional time costs in the detection process.

\noindent\textbf{The newly discovered RM-API misuses.}
Table~\ref{tab:diclosed_misuses} in the Appendix shows that \chatd{} discovers 115 security bugs on the applications integrated on six popular libraries, \revise{with 165 detected pairs and the details are shown in Figure~\ref{fig:type-pairs}.
More specifically, we categorize the RM-APIs based on the verbs extracted from their functionalities. The APIs labelled as ``other operations'' cannot have their corresponding actions directly extracted from the function names, such as \texttt{pcap\_list\_datalinks}. From the figure, it can be seen that APIs with the ``new'' operation account for as much as 18.79\%, making it the most prevalent. APIs with the ``malloc'' operation account for 10.3\%, ranking second. Other relatively common operations include ``create'', ``initiate'', ``open'', ``get'' and ``add'' have equal percentages of 5.45\%. In total, we identified 22 categories which can cover most of the resource-management APIs.
}

To further evaluate the security impact of these bugs, we follow the evaluation methods proposed by IPPO~\cite{liu2021detecting} and NDI~\cite{zhou2022non}. 
Three types of security bugs are found by \chatd{}: memory leak, use-after-free and double-free. 
95.7\% of the detected bugs would cause memory leaks and lead to memory exhaustion or Denial-of-Services(DoS) when triggered repeatedly. 
4 bugs would cause double-free and they could lead to memory corruption when triggered. 
One bug causes use-after-free and it may lead to DoS when triggered repeatedly.
We then use a fixed bug as a case study of a common security impact of the detected bugs. In \texttt{HandBrake} with 15.2k stars in GitHub, the function \texttt{av\_buffersrc\_parameters\_alloc()} allocates a new AVBufferSrcParameters instance and should be freed by the caller with \texttt{av\_free()} according to the documentation. However, the parameter \texttt{par} is double freed both before and within the error-handling structure. Therefore, it will lead to a double-free bug. If the bug is triggered, HandBrake may encounter memory corruption. This bug is detected by \chatd{} and the bug is confirmed and fixed by the software holder and the latest version is linked below\footnote{https://github.com/HandBrake/HandBrake/commit/ab467d13ef1f2a9a119c
2cd241c7c831ca207192}.

%%%%%%%%%%%%%%%%%%%%%%%%%%%%%%%%%%%%%%%%%%%%%%%%%%
%总体效果表格
\begin{table}[t]
\centering
\footnotesize
\setlength{\belowcaptionskip}{10pt}
\caption{Common RM-API pairs and bugs detected from four tools}
\label{tab:effect-3tools}
\begin{tabular}{>{\rule{0pt}{2.3ex}}m{1.2cm}m{0.4cm}m{0.4cm}
m{0.4cm}m{0.4cm}m{0.4cm}m{0.4cm}m{0.4cm}m{0.4cm}}
\hline
\multirow{2}*{\textbf{Library}} & 
\multicolumn{2}{c}{\textbf{Our work}} &
\multicolumn{2}{c}{\textbf{Advance}} &
\multicolumn{2}{c}{\textbf{HERO}} &
\multicolumn{2}{c}{\textbf{Goshawk}} \\
\cline{2-9}
 & \textbf{Pair} & \textbf{Bug} & \textbf{Pair} & \textbf{Bug} &  \textbf{Pair} & \textbf{Bug} & \textbf{Pair} & \textbf{Bug}\\
\hline
\textbf{Libevent} & \cellcolor{gray!25}\textbf{56} & 9& 4 & 3 & 1 & 0 &6&4  \\
\hline
\textbf{Libzip} & \cellcolor{gray!25}\textbf{23}& 2& 2 & 0 & 0 & 0 &0&0 \\
\hline
\textbf{Libexpat} & \cellcolor{gray!25}\textbf{6} & 0& 2 & 0 & 0 & 0 &0&0 \\
\hline
\textbf{Total} & \cellcolor{gray!25}\textbf{85} & 11 & 8 & 3  &  1 &  0&6&4\\
\hline
\end{tabular}
\vspace{-10pt}
\end{table}
%%%%%%%%%%%%%%%%%%%%%%%%%%%%%%%%%%%%%%%%%%%%%%%%%%

\subsection{RQ2: Comparison with the state-of-the-art}
%对比advance,hero和goshawk %并说明为什么不对比nlp-eye

To answer RQ2, we compare \chatd{} with the existing state-of-the-art vulnerability detection tools that utilize documentation analysis (Advance~\cite{lv2020rtfm}) and code analysis (HERO~\cite{wu2021understanding} and Goshawk~\cite{lyu2022goshawk}) for RM-API misuse detection. Table~\ref{tab:effect-3tools} presents the detection results of these four approaches on three popular libraries: Libevent~\cite{libevent}, Libexpat~\cite{libexpat} and Libzip~\cite{Libzip}.
In total, \chatd{} detects 85 RM-API pairs and 11 security bugs on the three libraries. 
The other three tools (Advance, HERO and Goshawk) can only detect 9.41\%, 1.17\% and 7.05\% of the RM-API pairs.
All of them can merely detect 27.27\%, 0\% and 36.36\% security bugs, respectively.

%\td{revise the following paragraph}

We further explore the reasons for the false negatives on RM-API misuse detection. Advance focuses on identifying sentences with strong sentiments, but most RM sentences do not have such tendencies, which leads to many misuses undiscovered. The difference between these two works is that \chatd{} can mine information beyond expert knowledge in identifying RM-API constraints.
Although Advance can detect various API usage norms, the current RM APIs do not contain most of them and introduce many false negatives.
HERO is designed to match functions accurately by utilizing the distinctive error-handling structure.
The results show that HERO has limitations in capturing a large number of library functions, resulting from two primary factors. 
Firstly, the applications exhibit the absence of involving library functions, posing a challenge in extracting relevant library API usage from the code. Secondly, HERO's emphasis on error-handling structures as a means to acquire API usages represents a unique approach not commonly employed in other software. 
%Collectively, these two reasons underscore the comparative limitations of HERO in contrast to \chatd{}.
Concerning Goshawk, designed to identify RM-API pairs through source code analysis, the lack of accurate API usage patterns can lead to significant false negatives in RM-API pairing and misuse detection. 
In this way, We found that \chatd{} outperforms other state-of-the-art tools in recognizing RM-API pairs and RM-API misuse detection.

%%The native CodeQL 
%%%%%%%%%%%%%%%%%-----CodeQL-----%%%%%%%%%%%%%%%%%%%%%%%%%
\begin{lstlisting}[language=c,frame=trBL,caption= {QL code for detecting the memory leak of function evconnlistener\_new\_bind },abovecaptionskip=10pt,belowcaptionskip=0pt,captionpos=b,label={lst:nativeQL}, keywordstyle=\bfseries\color{green!40!black},
commentstyle=\itshape\color{purple!40!black},
identifierstyle=\color{blue},
stringstyle=\color{orange},
basicstyle=\footnotesize\ttfamily,
breaklines=true,
numbers=left,
numbersep=-5pt,
stepnumber=1]
    import cpp
    from Function f, AllocationFunction alloc, FunctionCall call
    where call.getEnclosingFunction() = f and call.getTarget() = alloc
    select f, f.getName()
\end{lstlisting}
%%%%%%%%%%%%%%%%%%%%%%%%%%%%%%%%%%%%%%%%%%%%%%%%%%

%compare against native CodeQL
\revise{
Due to the static analysis tool CodeQL containing basic functionality for malloc API identification, we then compare \chatd{} against the basic functionality of CodeQL~\cite{NativeCodeQL} named  \texttt{AllocationFunction} and the QL code is shown in Listing~\ref{lst:nativeQL}. The results show that among all the six libraries, the basic function of CodeQL can only identify five allocation APIs which are \texttt{av\_realloc} from FFmpeg, \texttt{pcap\_list\_datalinks}, \texttt{pcap\_open\_dead\_with\_tstamp\_precision}, \texttt{pcap\_list\_tstamp\_types} and \texttt{pcap\_create} from Libpcap. 
The built-in malloc function identification functionality of CodeQL has difficulty in identifying most allocation APIs with multi-layered nesting.
\yy{The five identified allocation APIs directly call \texttt{malloc} which can be easily identified by CodeQL. However, most of the allocation APIs have at least two-layered nesting indirectly calling \texttt{malloc} and cannot be identified by CodeQL. For example, \texttt{evhttp\_connection\_base\_bufferevent\_new} from Libevent has allocation semantics and directly calls \texttt{mm\_malloc}. Although \texttt{mm\_malloc} directly calls \texttt{malloc} and represents clear allocation semantics, CodeQL cannot determine if \texttt{evhttp\_connection\_base\_bufferevent\_new} has allocation semantics. What's more, \texttt{zip\_source\_buffer} from Libzip is composed of \texttt{malloc} through at least three-layered nesting and \texttt{pcap\_findalldevs} from Libpcap is composed of \texttt{malloc} through at least four-layered nesting. Neither of them can be identified by CodeQL due to multi-layered nesting. The results show that the built-in malloc function identification of CodeQL conducts intra-procedural analysis which can only identify the allocation APIs directly calling \texttt{malloc} and leave a huge amount of RM-APIs with multi-layered nesting undiscovered} (72.7\% of the RM-APIs in GroundTruch have multi-layered nesting as illustrated in Section~\ref{sec:intro}). Therefore, the basic malloc function identification functionality of CodeQL is far less effective than \chatd{}, making it difficult to achieve the current task's objectives.}

\revise{
We also evaluate \chatd{} on the standard API-misuse dataset APIMU4C~\cite{gu2019empirical}, which contains 15 real-world RM-API misuses collected across three applications focusing on OpenSSL and basic C library.
\chatd{} can detect 14 of them and all of them have the same bug type ``memory leak'', the details are shown in Table~\ref{tab: details4data}. The missing case highlighted in grey focuses on \texttt{X509\_get\_ext\_d2i} which is deprecated with no direct detailed documentation and source code, therefore introduces false negatives of \chatd{}. 
}

%%%%%%%%%%%%%%%%%%%%%%%%%
% fillout the table with the details of misuses in APIMU4C
\begin{table}[h]
%\centering
%\vspace{10pt}
\caption{The RM-API misuses details from APIMU4C.}
\label{tab: details4data}
% \vspace{5pt}
\centering
\footnotesize
\setlength{\tabcolsep}{0.95mm}{
%\resizebox{0.40\textwidth}{!}{
\begin{tabular}{m{1.1cm}|m{0.8cm}|m{2.4cm}|m{3.7cm}}
%\begin{tabular}{c|c|c|c}
\hline
\textbf{Library}  & \textbf{Apps} & \textbf{API} & \textbf{Location}  \\
\hline
\cellcolor{gray!25}\textbf{OpenSSL} & \cellcolor{gray!25}\textbf{httpd} &\cellcolor{gray!25}\textbf{X509\_get\_ext\_d2i} &\cellcolor{gray!25}\textbf{httpd-.4.37/modules/ssl/ssl\_util \_ssl.c:209} \\
\hline
OpenSSL &httpd &BIO\_new  & httpd-.4.37/modules/ssl/ssl\_util \_ssl.c: 274 \\
\hline
OpenSSL &httpd &BIO\_new  & httpd-.4.37/modules/ssl/ssl\_util \_ocsp.c:361 \\
\hline
OpenSSL &curl & BIO\_new  & curl-curl-\_63\_0/lib/vtls/openssl.c:740 \\
\hline
Basic C &curl &strdup  & curl-curl-\_63\_0/lib/vtls/openssl.c:1442\\
\hline
OpenSSL &curl &BIO\_new  &curl-curl-7\_63\_0/lib/vtls/openssl.c:2997 \\
\hline
OpenSSL &curl &BIO\_new  & curl-curl-7\_63\_0/lib/vtls/openssl.c:3330\\
\hline
OpenSSL &curl &BIO\_new  & curl-curl-7\_63\_0/lib/vtls/openssl.c:3394 \\
\hline
Basic C &curl &malloc  &curl-curl-7\_63\_0/lib/escape.c:155 \\
\hline
Basic C &curl &strdup  & curl-curl-7\_63\_0/lib/smtp.c:518 \\
\hline
Basic C &curl &aprintf  &curl-curl-7\_63\_0/lib/smtp.c:531 \\
\hline
Basic C &curl &strdup  & curl-curl-7\_63\_0/lib/formdata.c:361 \\
\hline
OpenSSL &openssl &OPENSSL\_realloc  &crypto/asn1/f\_string.c:100 \\
\hline
OpenSSL &openssl &EC\_GROUP\_dup  &crypto/ec/ec\_ameth.c:300 \\
\hline
OpenSSL &openssl &EVP\_MD\_CTX\_new  &ssl/s3\_enc.c:44 \\
\hline
\end{tabular}
\vspace{-5pt}
}
\end{table}
%%%%%%%%%%%%%

%%%%%%%%%%%%%%%%%%%%%%%%%%%%%%%%%%%%%%%%%%%%%%%%%%
%ablation study
\begin{table}[t]
%\vspace{-10pt}
\centering
\small
\setlength{\belowcaptionskip}{10pt}
\caption{The Results of Ablation Study}
\label{tab:ablation-3tools}
%\begin{tabular}{>{\rule{0pt}{2.3ex}}m{1.8cm}|m{1.5cm}|m{1.5cm}|m{1.5cm}}
\begin{tabular}{c|c|c|c}
\hline
 & \textbf{Precision} & \textbf{Recall} & \textbf{F1 score}\\
\hline
\textbf{$Eval_0$} & \yy{32.14\%} & \yy{29.67\%} & \yy{30.86\%}   \\
\hline
\textbf{$Eval_1$} & \yy{77.97\%} & \yy{71.98\%} & \yy{74.9\%}  \\
\hline
\textbf{\chatd{}} & \yy{98.21\%} & \yy{90.65\%} & \yy{94.28\%}  \\
\hline
\end{tabular}
\vspace{-15pt}
\end{table}
%%%%%%%%%%%%%%%%%%%%%%%%%%%%%%%%%%%%%%%%%%%%%%%%%%

\subsection{RQ3: Ablation Study}
\label{sec:RQ3}

\revise{In this section, we conduct two experiments among \yy{six libraries (FFmpeg~\cite{Ffmpeg}, Libevent~\cite{libevent},
Libexpat~\cite{libexpat}, Libpcap~\cite{Libpcap}, Libzip~\cite{Libzip} and OpenLdap~\cite{Openldap})} to evaluate individual techniques. \chatd{} utilizes external knowledge, the Chain-of-Thought techniques and the cross-validation method for precisely RM-API pairing. Since the cross-validation should be implemented on the Chain-of-Thought results, we perform one evaluation for both of them. In the end, we evaluate the impact of different qualities of documentation for RM-API pairing. The details are shown as follows.   }

%在3个库（libevent,libzip和libexpat）上，尝试相同prompt，一个给定doc，一个不给doc，最终配对的RM-API数，进行对比
%\vspace{-5pt}
\paragraph{Contribution of external knowledge.}
\revise{
To figure out the contribution of external knowledge for RM-API pairing, We design two types of prompts for RM sentence identification: \textbf{Type-1:} containing no documentation (directly using LLMs, denoting $Eval_0$ in Table~\ref{tab:ablation-3tools}); \textbf{Type-2:} containing API documentation (denoting \chatd{} in the same table). We further perform Chain-of-Thought and cross-validation with the identified RM sentences to RM-API pairing. \yy{Among the six libraries, the precision rate of $Eval_0$ on RM-API pairing is 32.14\% and the F1 score is 30.86\%, which are not nearly as effective as \chatd{}.} The results fully explain the contribution of external knowledge to RM-API pairing. 
}

%在3个库（libevent,libzip和libexpat）上，尝试相同prompt，一个尝试思维链和交叉验证，一个直接给出releasing API，最终配对的RM-API数，进行对比
%\vspace{-8pt}
\paragraph{Contribution of Chain-of-Thought and cross-validation.}
\revise{
To evaluate the contribution of Chain-of-Thought and cross-validation, we also design two types of prompts: \textbf{Type-3:} prompting LLMs with external knowledge, denoting as $Eval_1$ in Table~\ref{tab:ablation-3tools}; \textbf{Type-4:} incorporating Chain-of-Thought and cross-validation with Type-3 prompting, denoting as \chatd{} in the same table. \yy{Performing Chain-of-Thought and cross-validation can obtain 25.94\% more RM-API pairs with an increase in precision from 77.97\% to 98.21\% and an increase in recall from 71.98\% to 90.65\%,} which demonstrates the contribution of Chain-of-Thought and cross-validation for RM-API pairing.
}
%%%%%%%%%%%%%%
%不同文档质量对RM-API匹配的影响
\begin{figure}[t]
    \centering
    \includegraphics[width=0.5\textwidth]{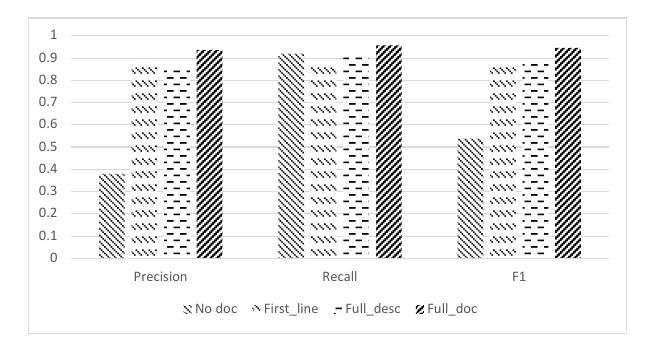}
    \caption{The impact of different qualities of documentation on RM-API pairing}
    %The comparison between different types of prompts and the method deployed in \chatd{}}
    \label{fig:data-ablation}
    \vspace{-5pt}
\end{figure}
%%%%%%%%%%%%%%%%%%%%%%%%%

%在3个库（libevent,libzip和libexpat）上，尝试相同sys prompt，不同文档质量以仅有一句话描述，有全部描述，有完整文档作为区分，对最终配对的RM-API数进行对比
%%我们认为无描述的文档是质量最低的，而仅有一句话描述的文档质量也明显低于有完整描述的文档。然而，从图xx中可以看出，是否有文档确实会对这一任务造成一定影响（F1 score有很大跃升），但是随着文档质量从一句话描述到有详细描述，F1分数的变化并不明显，且效果明显不如我们的工作。因此，我们得出结论我们的工作在无描述的API文档上有一定局限性，但是不会在其他质量类型的文档上有明显差异。
%\vspace{-20pt}
\paragraph{The impact of documentation quality.}
\revise{
To evaluate the impact of different qualities of documentation for RM-API pairing, we design \td{four} types of prompts representing different qualities and incorporate them with \chatd{} to perform the evaluation: 1) directly prompting \llm{} (referred \yy{to} as No\_doc); 2) using the prompt that includes the first sentence of the description (referred to as First\_line); 3) using the prompt that includes the entire description (referred to as Full\_desc); 4) and using the prompt that includes the entire document (referred to as Full\_doc). 
The comparative results are shown in Figure~\ref{fig:data-ablation}.
The results show that Full\_doc performs the best on the task among all the other prompts from three key dimensions: precision, recall and F1 score.
Specifically, we assume the API documentation with no description has the lowest quality and the API documentation with a one-sentence description also has significantly lower quality compared with the one including complete documentation. 
Figure~\ref{fig:data-ablation} shows that the presence of documentation has a certain impact on RM-API pairing, as the F1 score rapidly increases when the description appears. However, as the document quality improves from a one-sentence description to a detailed description, the change in the F1 score is not significant, and the performance is clearly not as good as Full\_doc which is also applied in our work (\chatd{}).
Therefore, we conclude that \chatd{} has limitations with undocumented API description, but does not show significant differences in other types of documentation quality.
}

\section{Explorations and Case study}
\label{sec:explore}

Although the comparisons show that \chatd{} outperforms the state-of-the-art API misuse detectors,
we are still curious about how \chatd{} achieves the task and to what extent it could be achieved.
In this section, we further analyze the similarities and differences between \chatd{} and the benchmark method (human inspection) through three research questions. 
Additionally, we explore the potential future research directions for subsequent research.

\ignore{
%%%%%%%%%%%%%%%%%%%%%%%%
\begin{figure}[t]
    \centering
    \includegraphics[width=0.45\textwidth]{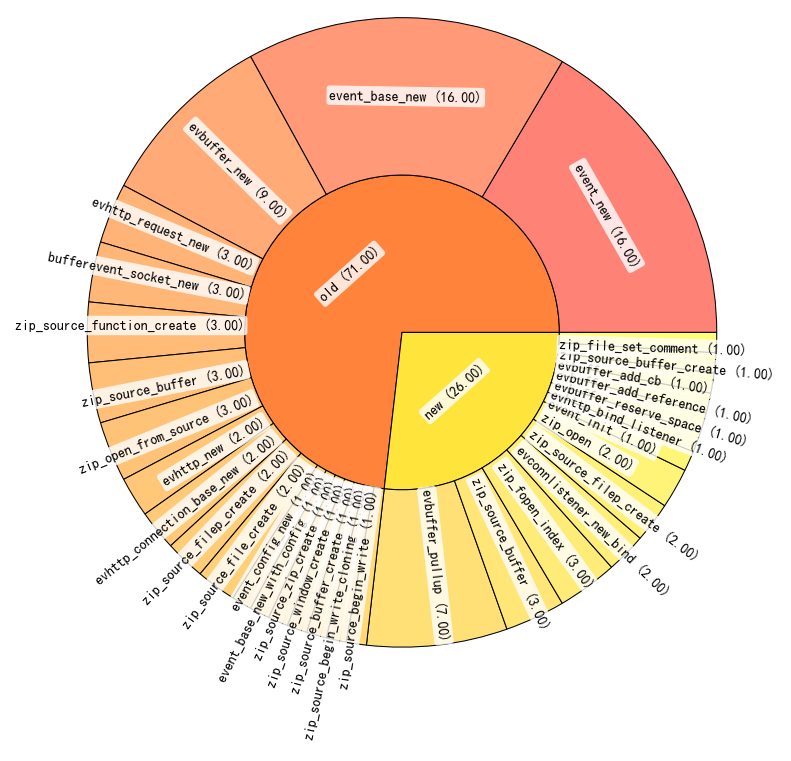}
    \caption{The distribution of API usage frequency}
    \label{fig:api-fre}
    \vspace{5pt}
\end{figure}
%%%%%%%%%%%%%%%%%%%%%%%%%
}

%human inspection
%%%%%%%%Framework%%%%%%%%%%
\begin{figure*}[t]
    \centering
    \includegraphics[width=0.8\textwidth]{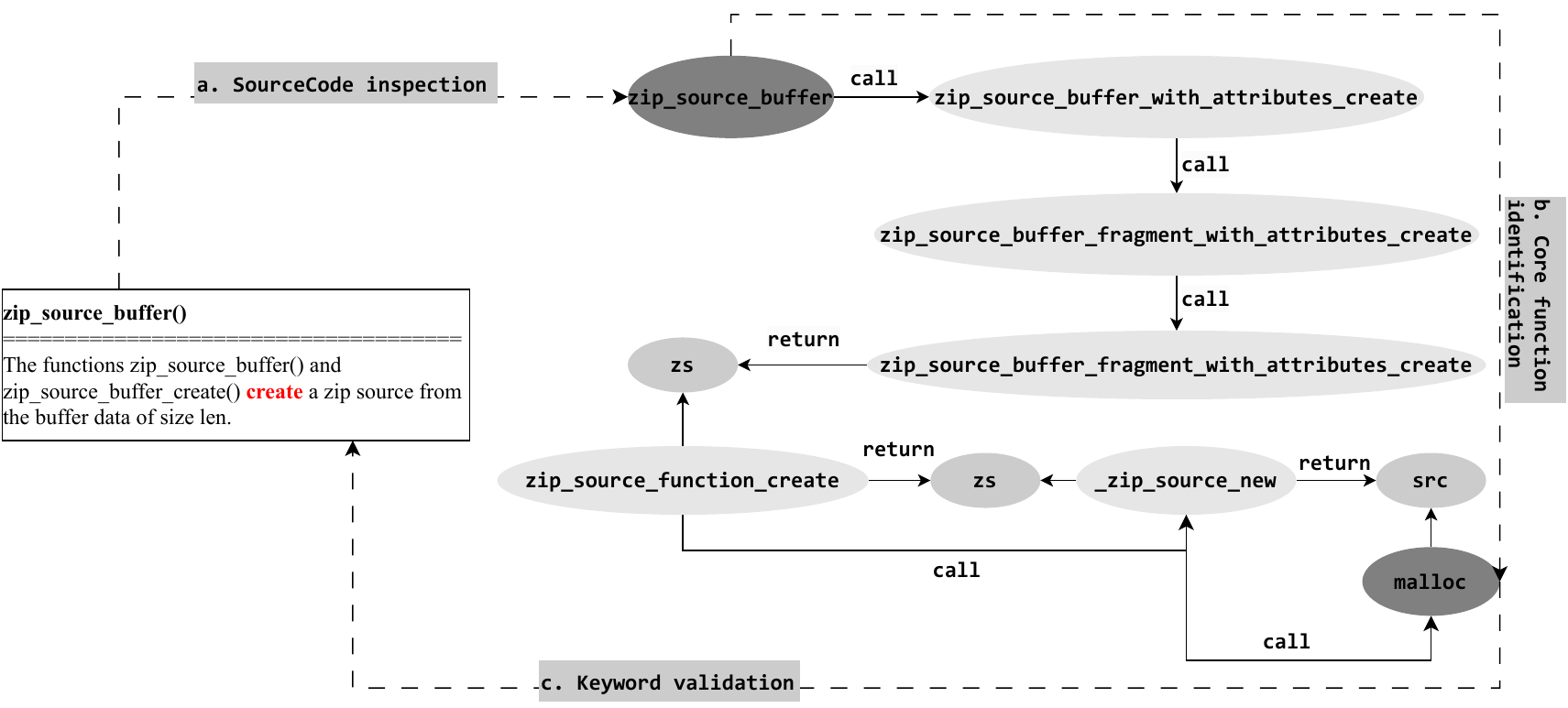}
    \caption{Manual Inspection of \texttt{zip\_source\_buffer}} 
    \label{fig:cross-src}
\end{figure*}
%%%%%%%%Framework%%%%%%%%%%

\noindent\textbf{Benchmark study.}
\label{para:pre-study}
To mimic the whole human inspections of RM-API pairs identification through documentation analysis during the software development, 
including RM-API identification and RM-API pairing based on the same RM-object type,
we chose three popular libraries for manual inspection evaluation: Libevent, Libzip and Libexpat.
The complete manual inspection is illustrated in Figure~\ref{fig:cross-src}. 
Take the function \texttt{zip\_source\_buffer} as an example, we first manually analyzed all the function calls within the function, through 6 iterations of function call analysis, we identified the core calling function \texttt{malloc} and confirmed the allocated object needs extra releasing operations, then we determined \texttt{zip\_source\_buffer} has allocation semantics, matching with the keyword ``create'' in the documentation.
Adopting this method on all the functions from three libraries, we found six commonly-used keywords were used for describing RM semantics, including ``allocate'', ``create'' and ``construct'' for resource allocation semantics; ``free'', ``release'' and ``deallocate'' for resource release semantics.
In total, \revise{the GroundTruth contains 92 malloc APIs and their corresponding releasing APIs from three libraries,} only 47 RM-API pairs can be identified \revise{through manual inspection.}

\begin{figure}[t]
    \centering
    \includegraphics[width=0.5\textwidth]{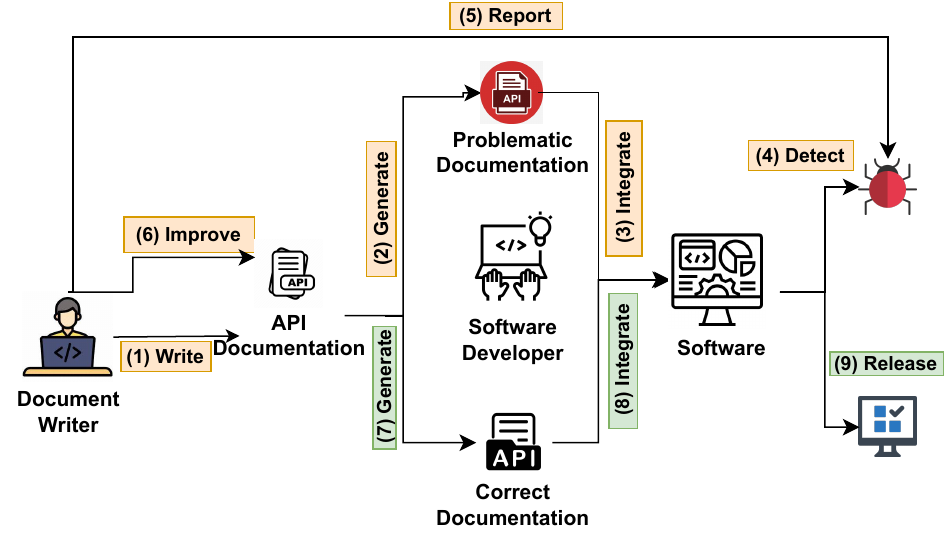}
    \caption{The software development cycle}
    \label{fig:doc-cycle}
    \vspace{-5pt}
\end{figure}

%\vspace{-5pt}
\noindent\textbf{Exploration\ding{182}: The relationship between buried API Constraints and the Documentation.}
\chatd{} performs better than human inspection in mining constraints from API documentation. By analyzing the usage frequency of both existing and newly discovered resource-allocation functions in software, the statistics show that the newly discovered functions are not widely used in the software. Some functions only exist once, originating from the function's source code rather than the calling code, indicating the rare usage identified in the development.
The main difference between the benchmark method and \chatd{} is that humans identify allocation functions using documentation keywords, excluding the code-analysis method due to the complexity of the code. 
%It aims to identify as many RM-API pairs as possible within the scope of expertise. 
The results show that \chatd{} excels in mining RM-API pairs, identifying 0.8 times more RM-API pairs than the benchmark method. 
However, due to the relatively low usage frequency of identified RM-API pairs, the number of RM-API misuses detected by \chatd{} doesn't significantly increase.
\chatd{} discovered 38 new RM-API pairs, but only 34.21\% of them are used, resulting in a total usage frequency of 36.63\% compared to the benchmark, leading to detecting fewer misuses.
Further, we infer the development cycle between the documentation writer, documentation and the users (referred to as software developers) through the analysis of the above phenomenon, as illustrated in Figure~\ref{fig:doc-cycle}. When functions are infrequently used or issues are not immediately found, documentation depends on regular updates. Incomplete documentation can cause misuse. Frequently used or security-issued functions often get better documentation, causing imbalance. This imbalance may leave developers unaware of security guidelines for lesser-used functions, leading to security issues.

%When functions have lower usage frequency or issues are not immediately discovered, the maintenance of documentation relies on regular updates by documentation writers. In cases where the documentation is incomplete in describing the function's functionality, it can lead to the misuse of the function.
%As more highly used functions or functions with detected security issues are reported to documentation developers, the documentation of the same library may become imbalanced in terms of quality. This imbalance may result in the descriptions for functions with lower usage frequency may also be incomplete. This can result in software developers being unaware of certain security guidelines when using those functions, leading to security problems.

%%%%%%%%%%%%%%
%软件开发流程
%%%%%%%%%%%%%%%%%%%%%%%%%

%\vspace{-5pt}
\noindent\textbf{Exploration\ding{183}: How does \chatd{} identify RM sentences?}
%\todo{add the description of table-3}
The statistical analysis of the identified RM-API pairs shows that \chatd{} introduces four resource-allocation APIs missing from identification due to the current method for extracting RM sentences.
%As shown in Table~\ref{tab:rm-sents-fn}, 
Most of the missed detections are attributed to the model itself. In the description, the RM sentence \textit{``The zip\_open\_from\_source() function opens a zip archive ...''} accurately describes the functionality of \texttt{zip\_open\_from\_source} and signifies that the current API has resource-allocation semantics. However, the model missed the identification and the misuse detection. Another reason for a missed detection is that the documentation itself lacks detailed descriptions, leaving the model without additional information to reference.

%\vspace{-5pt}
We conducted further analysis of RM sentences and found that the sentences containing keywords (as shown in the benchmark work, which is described in Section~\ref{para:pre-study}) constitute 52.17\% of them. The automation of the benchmark method for extracting RM sentences relies on matching corresponding keywords which may introduce both false positives and false negatives. However, thanks to \chatd{}'s comprehension abilities and the extensive training dataset, when identifying resource-allocation APIs, it goes beyond keyword-based recognition and can identify additional RM sentences. 
What's more, \chatd{} can detect RM sentences located in various parts of the documentation (e.g., description, first sentence, non-first sentence and return value)
%(as shown in Table~\ref{tab:rm-sents-exp}) 
compared to traditional matching methods.
The results reveal that leveraging the capabilities of LLMs could extend the boundaries of expert knowledge.

\begin{figure}[t]
    \centering
    \includegraphics[width=0.5\textwidth]{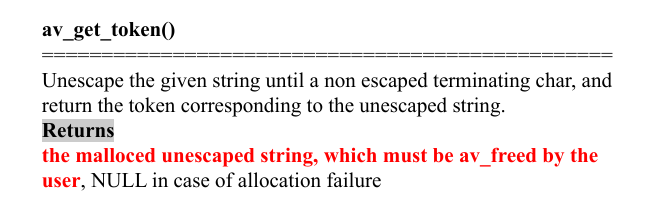}
    \caption{The documentation defect found by \chatd{}}
    \label{fig:incorrect-doc}
    \vspace{-5pt}
\end{figure}

\noindent\textbf{Exploration\ding{184}: How does \chatd{} identify RM APIs and paring them?}
As a generative model, \chat{} is trained on vast data and parameters, containing more expertise and a better ability to ``understand''.
In contrast to the benchmark method, which is time-consuming and labour-intensive, \chatd{} acquires more knowledge and a well-designed solution for getting accurate answers from LLMs.
For {\textit{RM-API Identification}}, people should understand the API descriptions and further identify the RM APIs.
But for \chatd{}, theoretically it should utilize the RM sentences as the evidence for RM-API identification. 
However, as a black-box model, it is hard to differentiate them from the training data.
Therefore, \chatd{} can identify more RM sentences than the benchmark method.  
For \textit{RM-API Pairing}, both the benchmark method and \chatd{} use a similar approach based on the same observation that resource-allocation functions and their corresponding releasing functions operate on objects of the same type. They achieve this by matching RM object descriptions with the function declarations of RM APIs to identify the RM object's type.
The key difference lies in how they determine the RM-object type. 
For the benchmark method, people tend to calculate the semantic similarity between RM object descriptions and the names of parameter/return value types to recognize the RM object type. In contrast, \chatd{} simplifies this step into one single prompt.
For most incomplete or loosely-organized descriptions, \chatd{} demonstrates stronger text-processing capabilities and can accurately identify RM-object types with relatively less information.
What's more, Figure~\ref{fig:incorrect-doc} shows a defect found by \chatd{} in the documentation. People are supposed to pair \texttt{av\_get\_token} with \texttt{av\_freed} according to the description, which means the caller of \texttt{av\_get\_token} is supposed to call \texttt{av\_freed} to prevent potential memory leak. However, the current presentation of the documentation is not only hard to follow, but also mentions the incorrect function, which does not exist. Such types of defects can result in incorrect constraints being extracted from existing work~\cite{lv2020rtfm}. However, \chatd{} is not only unaffected by incorrect information, but also capable of detecting such defects within the documentation.

\ignore{
%%%%%%%%%%%%%%%
%zip\_set\_archive\_comment sourcecode
\begin{lstlisting}[language=c,frame=trBL,caption= {zip\_set\_archive\_comment源码。},abovecaptionskip=10pt,belowcaptionskip=0pt,captionpos=b,label={listing: sourcecode-zip}, keywordstyle=\bfseries\color{green!40!black},
commentstyle=\itshape\color{purple!40!black},
identifierstyle=\color{blue},
stringstyle=\color{orange},
basicstyle=\footnotesize\ttfamily,
breaklines=true,
numbers=left,
numbersep=-5pt,
stepnumber=1]
    ZIP_EXTERN 
    int zip_set_archive_comment( zip_t *za, const char *comment, zip_uint16_t len) 
{
    zip_string_t *cstr;
    ...
    if (len > 0) {
	if (( cstr = _zip_string_new(( const zip_uint8_t *) comment, len, ZIP_FL_ENC_GUESS, &za->error)) == NULL)
	    return -1;
	if (_zip_guess_encoding(cstr, ZIP_ENCODING_UNKNOWN) == ZIP_ENCODING_CP437) 
        {
	    _zip_string_free(cstr);
	    zip_error_set(&za->error, ZIP_ER_INVAL, 0);
	    return -1;
	}
    }
    ...
    return 0;
} 
\end{lstlisting}
%%%%%%%%%%%%
}
%%%%%%%%%%%%%%%%%%%%%%%%%
%incorrect_doc found by chatdetector
%%%%%%%%%%%%%%%%%%%%%%%%%

%%%%%%%%%%%%%%
%zip_set 源码
\begin{figure}[t]
    \centering
    \includegraphics[width=0.4\textwidth]{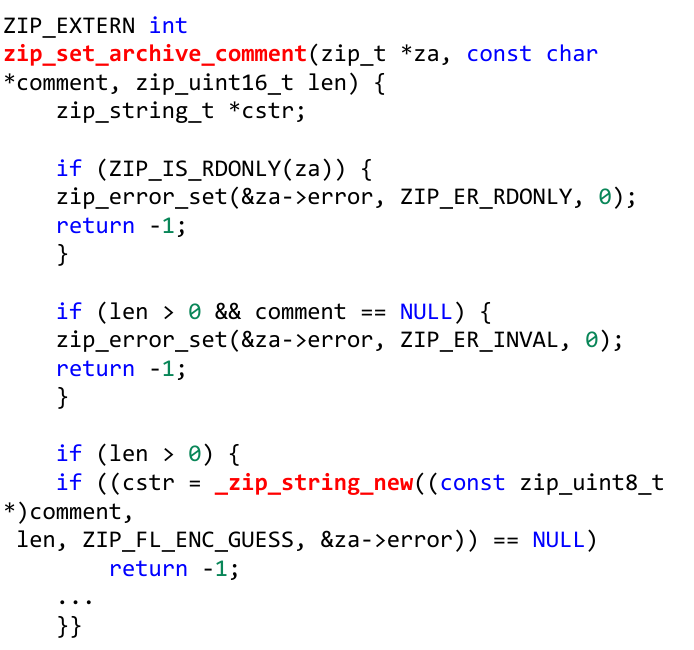}
    \caption{The source code of zip\_set\_archive\_comment}
    \label{fig:src-zip-set}
\end{figure}
%%%%%%%%%%%%%%%%%%%%%%%%%

%\vspace{-5pt}
\chatd{} uses the RM-object type provided in the prompt to determine the corresponding releasing function, while the benchmark method splits the procedure into RM object type identification and corresponding release API identification. 
In terms of results, \chatd{} can discover more RM-API pairs and achieve higher accuracy. However, since \chat{} is a black-box model and its training data is not accessible, it may produce judgments indicating resource allocation semantics even when there is no clear evidence of RM semantics in the text. For example, \chat{}'s analysis of the documentation for the function \texttt{zip\_set\_archive\_comment} led it to conclude that the function has resource allocation semantics.
Manual inspection of the source code in Figure~\ref{fig:src-zip-set} reveals that the function source code calls \texttt{\_zip\_string\_new} to allocate the resource of the string \textbf{cstr} and then promptly frees it by calling \texttt{\_zip\_string\_free} within the function.
Since this string is not passed as the return value or a variable to the function \texttt{zip\_set\_archive\_comment}, the function indeed has resource-allocation semantics but does not require further resource releasing.
Therefore, \chatd{}'s false positive for this function might be due to its internal dataset containing the source code of this function. When \chatd{} prompts \llm{} for an analysis of the function \texttt{zip\_set\_archive\_comment}, the response from \llm{} may also include an analysis of the source code, leading to \chatd{}'s false positive.
\vspace{-5pt}

\section{Discussions}
\label{sec:discuss}

%\todo{to be revised}
\noindent\textbf{The use of other techniques.}
As mentioned in Section~\ref{sec:design}, we apply the widely-used GPT-3.5-turbo model~\cite{gpt-3.5} to implement \chatd{}.
\revise{The total cost of six libraries using GPT-3.5-turbo is cheaper than 8.9 dollars and the cost of static program analysis is within a reasonable price.}
We also compared GPT-4.0-turbo and \chat{}, not only do the hallucinations still exist, but the cost of GPT-4.0-turbo is higher including the monthly user cost and the 20 times increased expense on each query.    
\revise{
Traditional type analysis~\cite{chen2020systematic, crix-security19} needs full source-code analysis of the function to acquire the type, while the LLM only takes one refined prompt for RM-object type identification which reaches nearly 93.55\% precision and the cost is cheaper than 0.00058 dollars per function on average. Therefore, utilizing LLMs exceeds the traditional type analysis in type identification.
}

%\vspace{-5pt}
\noindent\textbf{Limitations.}
\chatd{} introduces the novel concept of utilizing \llm{} to understand and validate its capability in analyzing official API documentation and extracting RM-API constraints beyond expertise for RM-API misuse detection. However, due to the black-box nature of \llm{}, it is challenging to understand how the model generates certain false positives and addresses them equally. 
\revise{
We also conducted a small comparison between CodeLLama-7b~\cite{codellama},vicuna-7b~\cite{vicuna}, GPT-3.5-turbo~\cite{gpt-3.5} and GPT-4.0~\cite{gpt-4.0} shows that hallucination still exists across different LLMs where directly using them will generate incorrect answers for RM-API identification. Considering their effectiveness and the corresponding costs, we chose GPT-3.5 to implement \chatd{}.
}
\chatd{} attempts to enhance the stability of \llm{}'s output inspired by ReAct framework, but this issue cannot be entirely eliminated, which remains further discussion.

%\vspace{-5pt}
\noindent\textbf{Future Work.}
With the increasing capabilities of \llm{}, the future of text understanding and analysis research has shifted from breaking through the bottlenecks of traditional NLP tools to exploring how to effectively leverage \llm{} to assist in other tasks. As tools make breakthrough advancements, tasks that were previously challenging due to tool limitations will present new opportunities. \revise{From one side, since the current explainable AI methods have not yet reached the field of LLMs~\cite{holzinger2022explainable,linardatos2020explainable,xu2019explainable}, the interpretability of LLMs could be a new direction worth exploring, which helps in exploring the potential of LLMs. }\revise{From the other side, }building upon the foundation of this work, the scope of extracted constraints can be gradually expanded, and further analysis of complex documentation can be conducted to enhance the work of API misuse detection.

\section{Related Work}\label{sec:relwork}

%\todo{too long for related work}

\noindent\textbf{Resource-Management API misuse detection.}
%\todo{revise}
Current static vulnerability detection methods can be divided into two types of approaches: code-based analysis and documentation-based analysis.
Previous research assumed API constraints were implemented in code and detected misuse by recovering constraints. 
Some works assume releasing resources post-allocation through analysis of error-handling structures~\cite{wu2021understanding,saha2013hector};
%Hector and HERO~\cite{wu2021understanding} . HERO identifies misuse from , while Hector extracts pairs from error-handling code. 
other works suggest mining frequent function usages to detect misuses~\cite{yun2016apisan,bian2020sinkfinder}.
%SinkFinder~\cite{} and APISan~\cite{yun2016apisan} suggest frequent function pairs represent software constraints. They detect deviations using frequent usage. 
However, code-based methods assume successful API rule implementation, leading to false negatives and positives due to pattern extraction.
Previous studies extracted API constraints from documentation to aid in detecting misuse. 
Some work identifies Resource Management (RM) APIs from Javadoc using resource-related keywords~\cite{zhong2011inferring}; other works involving templates matching within organized documentation~\cite{pandita2016icon} and extracting Integrated Assumptions (IAs) from loosely organized C documentation using sentiment-emphasized constraints~\cite{lv2020rtfm}.
Our work innovatively utilizes \llm{} for documentation understanding, adopting the inspirations from the ReAct framework, identifying the resource-allocation function, along with its resource-releasing function and constructing the RM-API constraints.

%Identifying Resource Management (RM) APIs typically involves analyzing API names and descriptions. Doc2Spec~\cite{zhong2011inferring} identifies RM function pairs from Javadoc using resource-related keywords. Pandita~\cite{pandita2016icon} suggests templates with subject-object relationships for C\# documentation. However, template matching faces challenges with disorganized documents. Advance~\cite{lv2020rtfm} extracts Integrated Assumptions (IAs) from loosely organized C documentation using sentiment-emphasized constraints.

%focuses on single-sentence rule extraction, different from previous research, by utilizing Question-Answer (QA) to understand the semantic content of sentences in library documentation, identify Resource Management (RM) APIs, and pair them, which is an approach not previously explored.

%\vspace{-5pt}
\noindent\textbf{Large Language Models in Security Research.}
In the era of LLMs' widespread adoption, research on their use as alternatives for addressing security issues has emerged. Some researchers propose to utilize LLMs for fuzzing: Deng et al~\cite{deng2023large-zero, deng2023large} proposed the direct utilization of \llm{} to generate input programs, enabling effective fuzz testing of deep learning libraries.
Another use is automated \revise{bug fixing} using \llm{}. Both Pearce et al.\cite{pearce2022examining} and Xia et al.\cite{xia2023automated,xia2023keep} propose LLM-based \revise{bug fixing} methods, confirming safe repairs without extra fine-tuning. 
However, the current research poses a blank space for utilizing \llm{} for API misuse detection.
Our work identifies concealed API usage patterns by comprehending the content of API documentation. It verifies the authenticity and accuracy of the answers through further processing and analysis of the reasoning process applied to \llm{}.

\ignore{
\noindent\textbf{Reasoning with Large Language Models.}
To get accurate answers from LLMs, Wei et al.~\cite{wei2022chain} proposed the \textit{Chain-of-Thought (CoT)} method, enabling complex reasoning capabilities through intermediate reasoning steps.
However, validating each potential reasoning path is time-consuming and incurs significant costs. Therefore, Wang et al.~\cite{wang2022self} and Yao et al.~\cite{yao2023tree} suggest replacing CoT's greedy decoding with the \textit{Self-Consistency} algorithm and enhancing CoT with the \textit{Tree-of-Thoughts} (ToT) framework, addressing traditional CoT issues.
%, known for superior mathematical performance. Further, Yao et al.~\cite{yao2023tree} enhance CoT with the \textit{Tree-of-Thoughts} (ToT) framework, treating reasoning as an intermediate step for common-sense question solving, addressing traditional CoT result verification issues.
Besides, ReAct framework~\cite{yao2023react} enables LLMs to interact with external tools to enhance answer reliability. Combining CoT with external knowledge, ReAct adopts a \textit{``Think-Act-Observe''} approach, decomposing tasks into plan execution, plan refinement with external knowledge, and final plan execution.
Consequently, \chatd{} does not make additional calls to online knowledge but rather uses API documentation as supplementary knowledge to the \llm{}, enabling the discovery and misuse detection of RM-API pairs.
}
\section{Conclusion}
\label{sec:conclusion}

In this paper, we introduce an automated RM-API misuse detector, \chatd{}, through task decomposition inspired by the ReAct framework to understand API official documentation with \llm{}, extracting RM-API constraints with a two-dimensional prompting for cross-validation and inconsistency-checking between the \llm{}' output and reasoning process, with which enables the real-world RM-API misuse detection.
In total, \chatd{} discovers 115 security issues, that could potentially leading severe security issues, and are ethically reported to the developers.
Compared to the state-of-the-art API misuse detectors, \chatd{} can additionally detect at least 80.85\% more RM-API pairs with a 98.21\% precision, demonstrating the ability of \llm{} on information retrieval beyond expertise. This work enlightens new research directions for API misuse detection aided by \llm{}, extending beyond overcoming the bottlenecks of traditional NLP tools to explore how to effectively leverage \llm{} for security research.

\section*{Acknowledgements}
We want to thank our shepherd and reviewers for their insightful and constructive comments which highly improve our paper.
The IIE authors are supported in part by CAS Project for Young Scientists in Basic Research (Grant No. YSBR-118), NSFC (92270204) and Youth Innovation Promotion Association CAS.

\vspace{10pt}

\bibliographystyle{IEEEtranS}
\bibliography{ref}

\section{Appendix}\label{sec:appendix}

\ignore{
\subsection{The difference between Chain-of-Thought and ReAct framework}
In Section~\ref{sec:backandstudy}, we introduce the basic concept of Chain-of-Thought~\cite{wei2023chainofthought} and ReAct framework~\cite{yao2023react}. Here, we show the difference between them through Figure~\ref{fig:diffincot}. 
%%%%exp-api-pairing%%%%%%
\begin{figure}[h]
    \centering
    \includegraphics[width=0.4\textwidth]{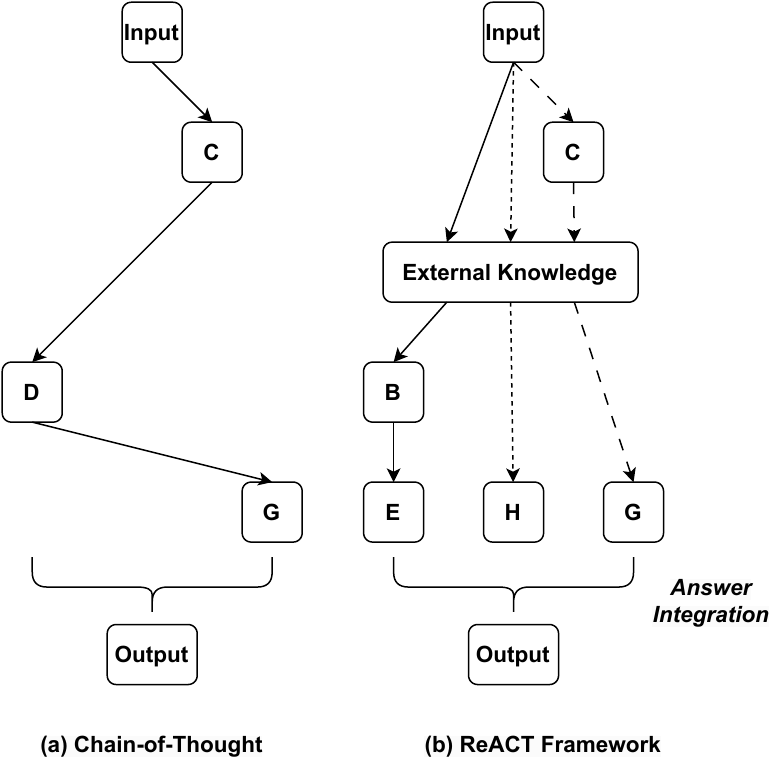}
    \caption{The difference between CoT and ReAct}
    \label{fig:diffincot}
    %\vspace{-5pt}
\end{figure}
%%%%%%%%%%%%%%%%%%%%%%%%%
}

\ignore{
\subsection{Initial attempt of LLMs}
In Section~\ref{sec:backandstudy}, we tried to prompt \chat{} for obtaining malloc API using the prompt shown in Figure~\ref{fig:init-m-api-iden} and for obtaining releasing API using the prompt shown in Figure~\ref{fig:direct-f-api}.
\vspace{-5pt}
%%%%exp-api-pairing%%%%%%
\begin{figure}[h]
    \centering
    \includegraphics[width=0.47\textwidth]{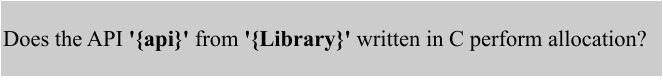}
    \caption{The prompt for malloc API identification without extra knowledge}
    \label{fig:init-m-api-iden}
    %\vspace{-5pt}
\end{figure}
%%%%%%%%%%%%%%%%%%%%%%%%%
%%%%exp-api-pairing%%%%%%
\begin{figure}[h]
    \centering
    \includegraphics[width=0.48\textwidth]{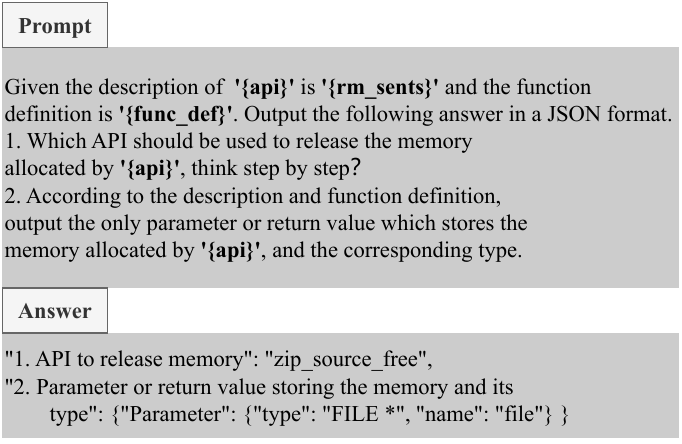}
    \caption{The prompt and answer for directly retrieving releasing API}
    \label{fig:direct-f-api}
    %vspace{-12pt}
\end{figure}
%%%%%%%%%%%%%%%%%%%%%%%%%
}

\subsection{Details of evaluation}

\begin{figure}[h]
    \centering
    \includegraphics[width=0.44\textwidth]{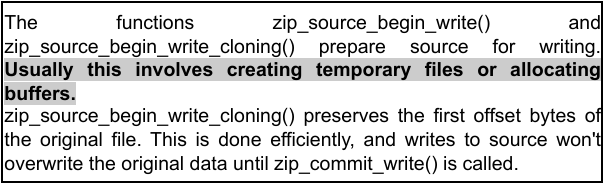}
    \caption{The documentation of zip\_source\_begin\_write \_cloning}
    \label{fig:zip-begin-doc}
    %\vspace{-10pt}
\end{figure}

\begin{figure}[h]
    \centering
    \includegraphics[width=0.44\textwidth]{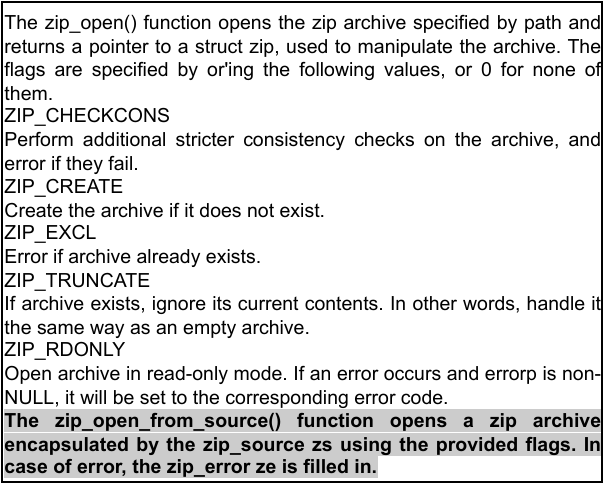}
    \caption{The documentation of zip\_open\_from\_source}
    \label{fig:zip-open-doc}
    %\vspace{-8pt}
\end{figure}

Several APIs are used for illustrating the effectiveness of \chatd{}, for example the documentation of \texttt{zip\_source\_begin\_write\_cloning} shown in Figure~\ref{fig:zip-begin-doc} and \texttt{zip\_open\_from\_source} shown in Figure~\ref{fig:zip-open-doc}.

%%%%%%%%%%%%%%%%%%%%%%%%%
%zip_source_begin_write_cloning的doc
%%%%%%%%%%%%%%%%%%%%%%%%%

%%%%%%%%%%%%%%%%%%%%%%%%%
%zip_open_from_source的doc
%%%%%%%%%%%%%%%%%%%%%%%%%

We choose six popular libraries for \chatd{} implementation and evaluation (Section~\ref{sec:evaluation}), the details are shown in Table~\ref{tab:lib-info}.
%%%%%%%%%%%%%%%%%%%%%%%%%%%%%%%%%%%%%%%%%%%%%%%%%%
%10 libs information
\begin{table}[h]
\centering
%\footnotesize
\small
\setlength{\belowcaptionskip}{10pt}
\caption{Library Information}
\label{tab:lib-info}
\begin{tabular}{m{1.2cm}|m{2.2cm}|m{1.2cm}|m{0.8cm}|m{0.9cm}}
\hline
\textbf{Library} & \textbf{Function Description} & \textbf{\#Line of Doc sentences} & \textbf{\#APIs} & \textbf{\#Apps}  \\
\hline

Ffmpeg & Multimedia & 3238 & 734 & 645\\
\hline
Openldap & Lightweight  Directory  Access Protocol & 5854 & 126 & 239 \\
\hline
Libevent & Asynchronous event notification & 2175 & 401 & 85 \\
\hline
Libpcap & Provides a portable framework for low-level network monitoring & 454 & 71 & 181 \\
\hline
Libzip & Reading, creating, and modifying zip archives & 2416 & 120 & 27 \\
\hline
Libexpat & Parsing XML & 422 & 70 & 275 \\
\hline
\textbf{Total} & / & \textbf{14,559} & \textbf{1,522} & / \\
\hline
\end{tabular}
%\flushright{\tabnote{*Applications}}
\vspace{-10pt}
\end{table}
%%%%%%%%%%%%%%%%%%%%%%%%%%%%%%%%%%%%%%%%%%%%%%%%%%

%\subsection{Examples of API documentation and source code}

%In total, 165 RM-API pairs are found by \chatd{} and we further categorize the types of detected pairs according to operations. 22 types are summarized and the details are shown in Figure~\ref{fig:type-pairs}.

\ignore{
%%%%%%%%%%%%%%
%findings in RM-API identification
\begin{figure}[T]
    \centering
    \includegraphics[width=0.4\textwidth]{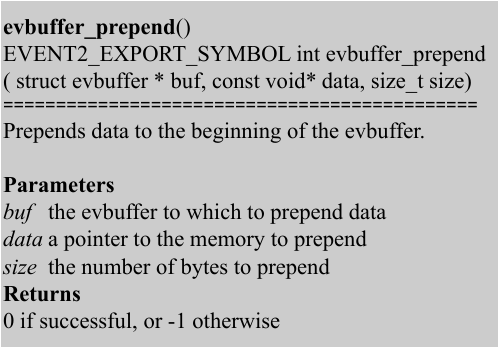}
    \caption{The documentation of evbuffer\_prepend.}
    \label{fig:doc-evbuffer}
    %\vspace{-5pt}
\end{figure}
%%%%%%%%%%%%%%%%%%%%%%%%%
}

%libevent缺少一个bug
%%%%%%%%%%%%%%%%%%%%
\begin{table*}[h]
\centering
%\small
%\footnotesize
\caption{Discovered RM-API misuses}
\label{tab:diclosed_misuses}
\begin{tabular}{>{\rule{0pt}{2.2ex}}m{1.1cm}<{\centering}|m{1.9cm}<{\centering}|m{4.2cm}<{\centering}|m{3.9cm}<{\centering}|m{2.0cm}<{\centering}|m{1.0cm}<{\centering}}
\hline
\textbf{Lib} & \textbf{Software} &  \textbf{Allocation API} & \textbf{Releasing API} & \textbf{Bug Type} & \textbf{\#Bugs} \\ \hline
\multirow{5}{*}{libevent} 
& seafile & event\_base\_new                            & event\_base\_free                       & memleak     & 1  \\
\cline{2-6}
&\multirow{4}{*}{evpp} & event\_base\_new                               & event\_base\_free                         & memleak     & 2                                  \\
        \cline{3-6}
        & & event\_config\_new                               & event\_config\_free                          & memleak     & 1                                  \\
        \cline{3-6}
        & & event\_new                               & event\_free                          & memleak     & 2                                  \\
        \cline{3-6}
        & & evhttp\_new                               & evhttp\_free                          & memleak     & 1                                  \\
    \cline{2-6}
    & transmission & event\_new                               & event\_free                          & memleak     & 1                                  \\\hline
\multirow{2}{*}{libzip}
& radare2 & zip\_source\_buffer\_create                               & zip\_source\_free                          & memleak     & 1                                  \\
    \cline{2-6}
    & OpenRCT2 & zip\_source\_buffer                               & zip\_source\_free                          & memleak     & 1                                  \\\hline
\multirow{8}{*}{ffmpeg} & \multirow{2}{*}{gpac} & av\_dict\_set                               & av\_dict\_free                          & memleak     & 2                                       \\
        \cline{3-6} 
        & & av\_dict\_copy                               & av\_dict\_free                          & memleak     & 1                                       \\
    \cline{2-6} 
    & \multirow{3}{*}{HandBrake} & av\_dict\_set                               & av\_dict\_free                          & memleak     & 21                         \\
        \cline{3-6} 
        & & avfilter\_graph\_create\_filter                               & avfilter\_free                          & memleak     & 1                            \\
        \cline{3-6} 
        & & av\_buffersrc\_parameters\_alloc                               & av\_free                          & double free     & 1                                      \\
        \cline{2-6}  &   \multirow{3}{*}{FFmpeg}  &av\_frame\_new\_side\_data& av\_frame\_remove\_side\_data & memory leak & 29\\
    \cline{3-6} 
    & & av\_frame\_alloc                               & av\_frame\_free                          & double free     & 1  \\
    \cline{3-6} 
    & & av\_new\_program                               & av\_free & memory leak     & 8                                     \\
    \cline{2-6} 
    & owntone-server & av\_dict\_set                               & av\_dict\_free                          & memleak     & 3                                      \\
    \cline{2-6}
    & vlc & av\_malloc                               & av\_freep                          & memleak     & 1                                      \\\hline
    \multirow{10}{*}{libpcap}
& \multirow{2}{*}{PF\_RING} & pcap\_findalldevs                               & pcap\_freealldevs                         & memleak     & 2                                  \\
        \cline{3-6}
        & & pcap\_compile                               & pcap\_freecode                          & memleak     & 4                                  \\
    \cline{2-6}
    & arp-scan & pcap\_compile                               & pcap\_freecode                          & memleak     & 1                                  \\
    \cline{2-6}
    & freeradius-server & pcap\_compile                               & pcap\_freecode                          & memleak     & 1                                  \\
    \cline{2-6}
    & nmap & pcap\_compile                               & pcap\_freecode                          & memleak     & 1                                  \\
    \cline{2-6}
    & ntopng & pcap\_compile                               & pcap\_freecode                          & memleak     & 2                                  \\
    \cline{2-6}
    & tcpdump & pcap\_compile                               & pcap\_freecode                          & memleak     & 1                                  \\
    \cline{2-6}
    & wireshark & pcap\_compile                               & pcap\_freecode                          & memleak     & 2                                  \\
    \cline{2-6}
    & \multirow{2}{*}{\shortstack{openvas-\\scanner}} & pcap\_compile                               & pcap\_freecode                          & memleak     & 1                                  \\
        \cline{3-6}
        & & pcap\_findalldevs                               & pcap\_freealldevs                          & memleak     & 2                                  \\\hline
\multirow{10}{*}{ldap} & freebsd-src & ldap\_search\_s                               & ldap\_msgfree                          & memleak     & 3 \\
    \cline{2-6}    & \multirow{2}{*}{\shortstack{freeradius-\\server}} & ldap\_result                               & ldap\_msgfree                          & double free     & 2                                      \\
        \cline{3-6}
        & & ldap\_result                               & ldap\_msgfree                          & memleak     & 1                                     \\
    \cline{2-6}
    & gpdb & ldap\_search\_s                               & ldap\_msgfree                          & memleak     & 1                                  \\
    \cline{2-6}
    & \multirow{4}{*}{openldap} & ldap\_search\_ext\_s                               & ldap\_msgfree                          & memleak     & 6                                 \\
        \cline{3-6}
        & & ldap\_url\_parse                               & ldap\_free\_urldesc                          & memleak     & 2                                  \\
        \cline{3-6}
        & & ldap\_initialize                               & ldap\_unbind\_ext                          & use after free     & 1                                  \\
        \cline{3-6}
        & & ldap\_initialize                               & ldap\_unbind\_ext                          & memleak     & 2                                  \\
    \cline{2-6}
    & yugabyte-db & ldap\_search\_s                               & ldap\_msgfree                          & memleak     & 1                                  \\
    \cline{2-6}
    & postgres & ldap\_search\_s                               & ldap\_msgfree                          & memleak     & 1                                  \\\hline
     
\multirow{1}{*}{Total}
& - & - & -  & -  & 115    \\
\bottomrule
\end{tabular}
\end{table*}

\end{document}